\documentclass[aps, showpacs,superscriptaddress,amsmath,amssymb,showkeys,floatfix]{revtex4-1}

\usepackage{graphicx}
\usepackage{dcolumn}
\usepackage{bm}
\usepackage[utf8]{inputenc}
\usepackage{braket}
\usepackage{amssymb,amsmath}
\usepackage[british]{babel}
\usepackage[autostyle=true]{csquotes}
\usepackage[colorlinks=true,citecolor=blue,linkcolor=blue,urlcolor=blue]{hyperref}
\usepackage{cleveref}
\crefname{equation}{Eq.\!}{Eqs.\!}
\crefname{figure}{Fig.\!}{Figs.\!}
\crefname{chapter}{Chap.\!}{Chaps.\!}
\crefname{section}{Sec.\!}{Secs.\!}
\crefname{appsec}{App.\!}{Apps.\!}
\usepackage[caption=false]{subfig}
\usepackage[export]{adjustbox}
\usepackage{wasysym}
\usepackage{graphicx}
\graphicspath{ {/} }
\newcommand{\twoDsite}[1]{\pmb{#1}}
\newcommand{\twoDsitesoft}[1]{\bm{#1}}
\newcommand{\ZS}[3]{\left. #1\right._{#3}^{#2}}
\newcommand{\cj}[1]{\left( #1 \right)^*}
\newcommand{\cjOnlySub}[1]{#1^*}
\newcommand{\hamilt}{\hat{H}}

\newcommand{\systemDomain}{\mathbb{S}}

\newcommand{\symmDomain}{\mathbb{D}^{\symmetryOp}}

\newcommand{\neighbourOf}[1]{\mathcal{N}(#1)}
\newcommand{\forallMacro}{\;, \; \forall \;}

\def\mystrut{\rule{0pt}{.4em}}
\newcommand{\strich}[1]{\overline{\mystrut#1}}
\newcommand{\symmetryOp}{\mathcal{S}}
\newcommand{\TS}[1]{\operatorname{\symmetryOp}_{#1}}

\newcommand{\TSInverse}[1]{\operatorname{\symmetryOp}_{#1}^{-1}}
\newcommand{\T}[2]{\operatorname{\symmetryOp}_{#1}( #2 )}

\newcommand{\TInverse}[2]{\inverse{\operatorname{\symmetryOp}_{#1}}\left( #2 \right)}
\newcommand{\absolute}[1]{\left| #1\right|}
\newcommand{\inverse}[1]{#1^{-1}}

\newcommand{\qWithoutI}[4]{q_{#3}^{#4}}

\newcommand{\qPlusWithoutI}[3]{\qWithoutI{}{#1}{#2}{#3}}

\newcommand{\closedLoopDomain}{\mathcal{C}_{\pmb{\ocircle}}}

\let\originalleft\left
\let\originalright\right
\renewcommand{\left}{\mathopen{}\mathclose\bgroup\originalleft}
\renewcommand{\right}{\aftergroup\egroup\originalright}

\crefname{secinapp}{appendix}{appendices}
\Crefname{secinapp}{Appendix}{Appendices}
\begin{document}

\title{Non-Local Currents and the Structure of Eigenstates in Planar Discrete Systems with Local Symmetries}

\author{M. Röntgen}
\email[]{mroentge@physnet.uni-hamburg.de}
\affiliation{%
Zentrum für optische Quantentechnologien, Universität Hamburg, Luruper Chaussee 149, 22761 Hamburg, Germany
}%
\author{C. V. Morfonios}%
\email[]{christian.morfonios@physnet.uni-hamburg.de}
\affiliation{%
	Zentrum für optische Quantentechnologien, Universität Hamburg, Luruper Chaussee 149, 22761 Hamburg, Germany
}%
\author{F.K. Diakonos}%
\email[]{fdiakono@phys.uoa.gr}
\affiliation{%
	Department of Physics, University of Athens, GR-15771 Athens, Greece
}%
\author{P. Schmelcher}
\email[]{pschmelc@physnet.uni-hamburg.de}
\affiliation{%
	Zentrum für optische Quantentechnologien, Universität Hamburg, Luruper Chaussee 149, 22761 Hamburg, Germany
}%
\affiliation{%
	The Hamburg Centre for Ultrafast Imaging, Universität Hamburg, Luruper Chaussee 149, 22761 Hamburg, Germany
}%

\date{\today}

\begin{abstract}
	Local symmetries are spatial symmetries present in a subdomain of a complex system. By using and extending a framework of so-called non-local currents that has been established recently, we show that one can gain knowledge about the structure of eigenstates in locally symmetric setups through a Kirchhoff-type law for the non-local currents. The framework is applicable to all discrete planar Schrödinger setups, including those with non-uniform connectivity. Conditions for spatially constant non-local currents are derived and we explore two types of locally symmetric subsystems in detail, closed-loops and one-dimensional open ended chains. We find these systems to support locally similar or even locally symmetric eigenstates.
\end{abstract} 

\maketitle
\section{\label{sec:Introduction}Introduction}
Local symmetries, i.e. symmetries that are only present in a spatial part of a given system, are ubiquitous in  nature, a popular example being  quasicrystals \cite{PhysRevLett.53.2477, PhysRevLett.53.1951,PhysRevLett.63.310, LIFSHITZ1996633,Morfonios2014}. Due to the long-range order of quasicrystals, one can always find structures of equal structure which can be described by local symmetries. Other examples are large molecules \cite{doi:10.1021/jp011482j,Domagala:wf5030} and, in general, systems where the global symmetry is broken due to defects. Beyond this, a second class of systems are those which are specifically designed in such a way that they possess local symmetries. Examples therefore are photonic multilayers \cite{PhysRevLett.55.1768,Aissaoui,PhysRevA.81.053808,:/content/aip/journal/apl/80/17/10.1063/1.1468895} or photonic waveguide arrays \cite{Davis:96,:/content/aip/journal/apl/71/23/10.1063/1.120327,Naether:12,PhysRevA.75.053814}.

Despite their widespread presence in both natural and artificial physical systems, a systematic and in-depth treatment of the influence of local symmetries on a system's behaviour is still missing. A reason for this might lie in the tools used.
In quantum systems, for example, the treatment of symmetries is based on the determination of the Hamiltonian group, i.e. the set of operators commuting with the Hamiltonian of the considered system. These operators usually refer to global symmetry transforms such as translation or reflection. The corresponding Hamiltonian eigenstates are also eigenstates of the irreducible representations of the symmetry operators, leading
to states with definite parity (reflection) \cite{Zettili200902} or Bloch-states (discrete translation) \cite{BlochTheorem}. For operators $\hat{\Sigma}_{L}$ describing local symmetries valid only in a limited spatial domain, however, we have $[\hat{H},\hat{\Sigma}_{L}] \ne 0$ in general. Does this mean that local symmetries do not affect the eigenstates of the Hamiltonian? Or is it possible to gain additional knowledge about the structure of eigenstates in locally symmetric systems using other means?

Recently a framework for the treatment of local symmetries in one-dimensional discrete setups has been established \cite{2016arXiv160706577M}, motivated by corresponding results for discrete local symmetries in continuous one-dimensional systems \cite{PhysRevA.87.032113,PhysRevA.88.033857,PhysRevLett.113.050403,PhysRevB.92.014303}. The very spirit of this framework is to use local symmetries to define new quantities, so-called non-local currents obeying suitably defined continuity equations.
For eigenstates in one-dimensional discrete systems, the non-local currents have two interesting properties: Firstly, for a given site the sum of a source term and the two ingoing non-local currents vanishes. This is a Kirchhoff-type law, and because it contains non-local currents, we call it a \emph{non-local} Kirchhoff law. Secondly, the non-local currents are piecewise constant throughout the corresponding domains of local symmetry, no matter how the wavefunction looks like, and thus this constancy may be used to derive a first insight into the structure of the system's eigenstates.

The present paper is mainly motivated by two questions: (i) How can one generalize the above framework to two-dimensional setups and (ii) are there structural properties of eigenstates that may be derived using non-local currents? Considering the first question, two important differences between one and higher discrete dimensions must be considered. The first difference is related to the possible number of different local symmetries. In one dimension there are only reflection and translation symmetries on the line. In higher dimensions, discrete rotations, plane reflections and even more general site permutations are possible. The second difference is related to the possible non-uniform connectivity across a system. That is, sites at different locations may have differing numbers of neighbours. Examples of this kind are Lieb lattices \cite{PhysRevLett.62.1201, PhysRevLett.114.245503} or in general all systems where the number of neighbours can vary from site to site.
In this work we focus on eigenstates and extend the framework of non-local currents to planar setups.
In contrast to one-dimensional systems, the non-local Kirchhoff law for stationary states here generally includes more than two currents for a given site, depending on its connectivity. This means that the non-local currents are no longer constant, even within domains of local symmetries. However, we give conditions that enable one to derive a summed non-local current which is, even in planar systems, constant throughout a domain of local symmetry. Furthermore, we incorporate the treatment of setups with non-uniform connectivity into the framework and investigate the influence of both the mapping and the connectivity of a setup onto its non-local currents.

We explore the consequences of higher and varying connectivity on the stationary non-local currents of two classes of setups: open-ended chains and closed loops, both being present in \cref{fig:localSymmetricSystem}. For the first class, we find eigenstates that are locally similar. Here and in the following, a \emph{locally similar eigenstate} denotes an eigenstate where some of the constituent amplitudes are related to each other by an eigenvalue-dependent constant. If this constant is $\pm1$, then the eigenstate is \emph{locally symmetric} or \emph{locally anti-symmetric}, respectively. For the second class, our results can be classified according to the local symmetry of the closed loop. For a local reflection symmetry, the eigenstates are locally symmetric and anti-symmetric. For local translation symmetry, \emph{some of the} eigenstates are locally symmetric.
Using non-local currents, the above properties can easily be proven and interpreted. This shows that the framework of non-local currents can be used as a practical tool and paves the way towards understanding the behaviour of complex systems on a local level.

This work is organized as follows. \Cref{sec:setup,sec:localSymmetries} describe the setup and introduce the necessary symmetry mappings needed to describe local symmetries. In \cref{sec:oneD} we introduce non-local currents and list some of their properties, followed by an in-depth graphical description and analysis of three example systems in \cref{sec:domainsOfLocalSymmetriesAndSourceTerms}. \Cref{sec:PermutationMapping} investigates the different possible types of local symmetries. After deriving constant summed currents in \cref{sec:SummedCurrents}, we will finally demonstrate two example systems featuring locally symmetric eigenstates in \cref{sec:exampleSystems}. Following the conclusions, the proofs for the various statements of this paper are then presented in \cref{sec:ProofConstancy,sec:SummedCurrentsProof,sec:ProofOpenEnd,sec:ProofClosedLoop}.

\section{Theoretical Framework}
\subsection{Setup and notation} \label{sec:setup}
We represent the Schrödinger equation (setting $\hbar = 1$)
\begin{equation} \label{eq:Schroedinger}
i\partial_t \ket{\Psi} = \hamilt \ket{\Psi}
\end{equation}
in a discrete basis set $\{\ket{\twoDsite{n}} \}$ where $\ket{\twoDsite{n}}$ is a localized excitation on site $\twoDsite{n}$. In the following we restrict ourselves to a tight-binding system, i.e. the hopping $h_{\twoDsite{n},\twoDsite{m}}$ between sites $\twoDsite{n}$ and $\twoDsite{m}$ vanishes whenever $\twoDsite{n}$ and $\twoDsite{m}$ are no neighbours.
The non-interacting Hamiltonian elements can then be written as
\begin{equation} \label{eq:discreteHamiltonian}
\braket{\twoDsite{n}|\hamilt|\twoDsite{n'}} = \begin{cases} v_{\twoDsite{n}}, & \mbox{if } \twoDsite{n'} = \twoDsite{n} \\ h_{\twoDsite{n},\twoDsite{n'}} \ne 0, & \twoDsite{n'} \in \neighbourOf{\twoDsite{n}}  \\ 0, & \mbox{else}\end{cases}
\end{equation}
where $v_{\twoDsite{n}}$ is the on-site potential at site $\twoDsite{n}$ and $\neighbourOf{\twoDsite{n}}$ denotes the set of neighbours of site $\twoDsite{n}$. 
In order to concentrate on the main features of locally symmetric planar setups, in this paper we consider only horizontal and vertical next-neighbour hoppings, as can be seen in \cref{fig:localSymmetricSystem}.
\subsection{Mappings for the description of local symmetries in tight-binding systems} \label{sec:localSymmetries}
Let us now introduce local symmetries. Simply speaking, a local symmetry is the invariance of \emph{some} on-site potentials $v_{\twoDsite{n}}$ and hoppings $h_{\twoDsite{n},\twoDsite{m}}$ of the system under the bijective transformations of a certain subset of sites and links. Here, the transformation acts on sites and links since they are the fundamental quantities on which on-site potentials and hopping terms are defined. We can write the underlying coordinate transformations as $\TS{\systemDomain}: \systemDomain \rightarrow \systemDomain$, i.e. as a bijective mapping from site $\twoDsite{n} \in \systemDomain$ to another (or the same) site $\strich{\twoDsite{n}} =: \T{\systemDomain}{\twoDsite{n}} \in \systemDomain$, where $\systemDomain$ denotes the whole system. Note that we have chosen $\systemDomain$ as the domain of $\TS{\systemDomain}$ in order to allow a later generalisation to global symmetries as well. The above statement of a local invariance of on-site potentials and hopping terms then transfers to the equation
\begin{equation} \label{eq:locSymmetryCondition}
H_{\twoDsite{m},\twoDsite{n}} = H_{\strich{\twoDsite{m}},\strich{\twoDsite{n}}} \forallMacro \twoDsite{m},\twoDsite{n} \in \symmDomain
\end{equation}
where $\symmDomain$ denotes the domain of local symmetry w.r.t. the mapping $\TS{\systemDomain}$. As noted above, this formulation enables us to describe global symmetries as well, for which $\symmDomain = \systemDomain$. For local symmetries, however, we usually have $\symmDomain \subset \systemDomain$.
Note that a system may feature more than one local symmetry, and  different local symmetries may overlap. We will investigate this issue in more detail in \cref{sec:domainsOfLocalSymmetriesAndSourceTerms} of this paper.
\begin{figure}[tbp]
	\centering
	\includegraphics[max size={\linewidth}{0.8\textheight}]{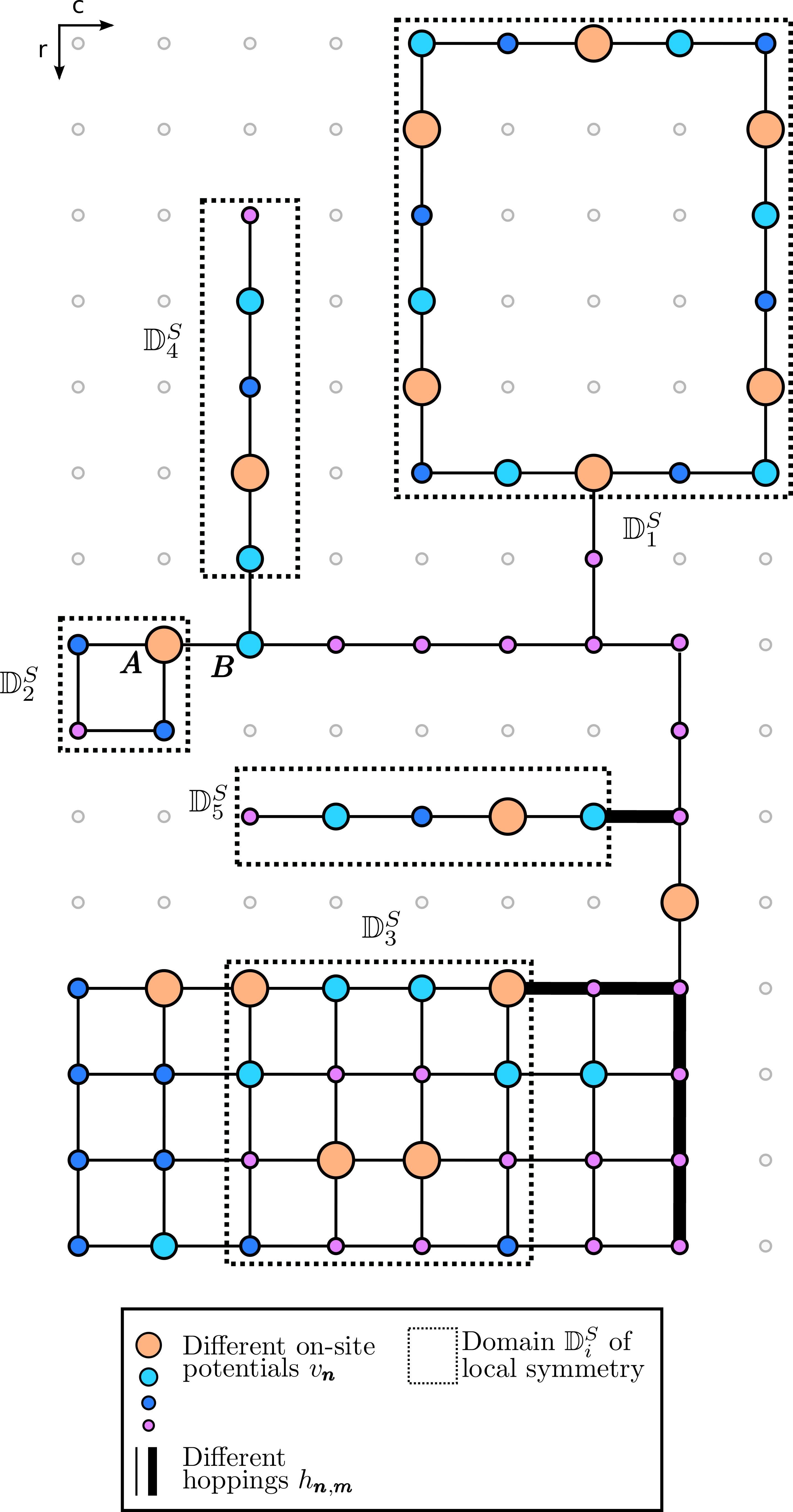}
	\caption{A system possessing local symmetries, with five domains $\symmDomain_{1}\ldots \symmDomain_{5}$ of local symmetries: clockwise or anti-clockwise translation ($\symmDomain_{1}$), diagonal ($\symmDomain_{2}$) and vertical ($\symmDomain_{3}$) reflection, and combinations of discrete rotations and translations ($\symmDomain_{4}$ and $\symmDomain_{5}$). Note that the mapping $\TS{\systemDomain}$ used to describe these local symmetries is not shown here.}
	\label{fig:localSymmetricSystem}
\end{figure}

Inspecting \cref{fig:localSymmetricSystem}, one realizes immediately that the considered system possesses local symmetries. These are clockwise and anti-clockwise translations ($\symmDomain_{1}$), reflection at the diagonal ($\symmDomain_{2}$) and the vertical ($\symmDomain_{3}$) centre line and, lastly, a combination of two different operations for the two closely related domains $\symmDomain_{4}$ and $\symmDomain_{5}$. For $\symmDomain_{4}$, this combination is given by an anti-clockwise rotation around its dark blue central site by $90^{\circ}$ followed by a translation of $(2,5)$ sites to the lower right. For $\symmDomain_{5}$, this combination is given by a clockwise rotation around its dark blue central site by $90^{\circ}$ followed by a translation of $(-2,-5)$ sites to the upper left.
Note that our above definition of a local symmetry does only take hopping terms \emph{within} $\symmDomain$ into account, as can clearly be seen at the first domain $\symmDomain_{1}$. In the course of this work, we will investigate \cref{fig:localSymmetricSystem} in more detail and see how some of its parts enable us to design eigenstates in such a way that they clearly exhibit the effects of local symmetries.
\subsection{Non-local currents and their properties\label{sec:oneD}}
In general the operator $\hat{\Sigma}_{L}$ describing local symmetries does not commute with the system's Hamiltonian, i.e. $[\hamilt,\hat{\Sigma}_{L}] \ne 0$. Therefore, the eigenstates of $\hat{\Sigma}_{L}$ are in general not eigenstates of $\hamilt$, contrary to global symmetries. Thus, the effects of local symmetries are not \enquote{visible} by means of the commutator $[\hamilt,\hat{\Sigma}_{L}]$ and the question arises whether there are other tools to investigate the effects of local symmetries. Recently, a new framework for the treatment of local symmetries in one-dimensional discrete systems has been achieved in \cite{2016arXiv160706577M} by the formulation of a non-local continuity equation for one-dimensional arrays. In the following, we generalize this non-local continuity equation for a general multidimensional system with multiple site connectivity. Basis of our approach is a non-local density operator $\hat{\sigma}_{\twoDsite{n}} = \ket{\twoDsite{n}}\bra{\strich{\twoDsite{n}}}$ constructed from site $\twoDsite{n}$ and its transformed counterpart $\strich{\twoDsite{n}} = \T{\systemDomain}{\twoDsite{n}}$. Using the general wavefunction $\ket{\Psi} = \sum_{\twoDsite{n}} \Psi_{\twoDsite{n}} \ket{\twoDsite{n}}$, the expectation value of this operator evaluates as
\begin{equation*}
\sigma_{\twoDsite{n}} = \braket{\Psi | \hat{\sigma}_{\twoDsite{n}} | \Psi} = \sum_{\twoDsite{k},\twoDsite{k}'} \Psi_{\twoDsite{k}}^{*} \braket{\twoDsite{k} | \twoDsite{n}} \braket{\strich{\twoDsite{n}}| \twoDsite{k}'} \Psi_{\twoDsite{n}'} = \Psi_{\twoDsite{n}}^{*} \Psi_{\strich{\twoDsite{n}}} .
\end{equation*}
By using the Schrödinger \cref{eq:Schroedinger}, we can evaluate the time derivative of $\hat{\Sigma}_{\twoDsite{n}}$, getting \begin{equation} \label{eq:nonLocalCont}
\partial_{t} \sigma_{\twoDsite{n}} = q_{\twoDsite{n}} -i(v_{\strich{\twoDsite{n}}} - v_{\twoDsite{n}}^{*}) \sigma_{\twoDsite{n}} = q_{\twoDsite{n}} - i\beta_{\twoDsite{n}} \sigma_{\twoDsite{n}}
\end{equation}
where $\beta_{\twoDsite{n}} = (v_{\strich{\twoDsite{n}}} - v_{\twoDsite{n}}^{*})$ and 
\begin{equation} \label{eq:definitionOfQn}
i\cdot q_{\twoDsite{n}} = \sum_{\twoDsite{m} \in \neighbourOf{\strich{\twoDsite{n}}}} h_{\strich{\twoDsite{n}},\twoDsite{m}} \cjOnlySub{\Psi_{\twoDsite{n}}} \Psi_{\twoDsite{m}} - \sum_{\twoDsite{m}\in \neighbourOf{\twoDsite{n}}} h_{\twoDsite{n},\twoDsite{m}}^{*} \Psi_{\strich{\twoDsite{n}}} \cjOnlySub{\Psi_{\twoDsite{m}}}
\end{equation}
with $\neighbourOf{\twoDsite{n}}, \neighbourOf{\strich{\twoDsite{n}}}$ denoting the set of neighbours of site $\twoDsite{n},\strich{\twoDsite{n}}$, respectively. Before we proceed, let us briefly comment on the number of summands occurring in \cref{eq:definitionOfQn}. As there is one term for each neighbour of $\twoDsite{n}$ and $\strich{\twoDsite{n}}$, respectively, there are $\absolute{\neighbourOf{\strich{\twoDsite{n}}}}$ summands in the left and $\absolute{\neighbourOf{\twoDsite{n}}}$  in the right sum, with $\absolute{\neighbourOf{\twoDsite{m}}}$ denoting the total number of neighbours of site $\twoDsite{m}$. It is important to stress that the two sums may have \emph{different} numbers of summands, i.e. $\absolute{\neighbourOf{\twoDsite{n}}}$ and $\absolute{\neighbourOf{\strich{\twoDsite{n}}}}$ \emph{do not need to be equal}. This is the case whenever $\twoDsite{n}$ and its mapped counterpart $\strich{\twoDsite{n}}$ have different connectivities.

We will now introduce non-local currents $\qPlusWithoutI{}{\twoDsite{n},\twoDsite{m}}{}$ and motivate their naming. After a short discussion of their properties we will use $\qPlusWithoutI{}{\twoDsite{n},\twoDsite{m}}{}$ to rewrite \cref{eq:nonLocalCont} which will result in the \emph{non-local continuity equation}. This equation will then prove to be the main tool for the treatment of systems possessing local symmetries, although it is applicable to all discrete planar systems, including the ones without local symmetries.

We define the non-local current flowing from site $\twoDsite{n}$ to $\twoDsite{m}$ as
\begin{equation} \label{eq:definitionOfq}
\qPlusWithoutI{}{\twoDsite{n},\twoDsite{m}}{}:= \frac{1}{i} \cdot \big(\ZS{h}{}{\twoDsite{\strich{n}},\twoDsite{\strich{m}}} \Psi_{\twoDsite{n}}^{*} \ZS{\Psi}{}{\strich{\twoDsite{m}}} -  \ZS{h}{*}{\twoDsite{n},\twoDsite{m}} \Psi_{\strich{\twoDsite{n}}} \Psi_{\twoDsite{m}}^{*}\big) .
\end{equation}
To motivate naming $\qPlusWithoutI{}{\twoDsite{n},\twoDsite{m}}{}$ a non-local current, let us note that one can derive the well-known probability current
\begin{equation} \label{eq:usualCurrent}
j_{\twoDsite{n},\twoDsite{m}} = \frac{1}{i} \cdot \big( \ZS{h}{}{\twoDsite{n},\twoDsite{m}} \cjOnlySub{\Psi_{\twoDsite{n}}} \Psi_{\twoDsite{m}} -\ZS{h}{*}{\twoDsite{n},\twoDsite{m}} \Psi_{\twoDsite{n}} \cjOnlySub{\Psi_{\twoDsite{m}}} \big)
\end{equation}
(where $\twoDsite{n},\twoDsite{m}$ are only restricted to be elements of $\systemDomain$, i.e. elements of the set of all sites of the system) from $\qPlusWithoutI{}{\twoDsite{n},\twoDsite{m}}{}$ by setting  $\strich{\twoDsite{n}} = \twoDsite{n} \forallMacro \twoDsite{n} \in \systemDomain$. Therefore, $\qPlusWithoutI{}{\twoDsite{n},\twoDsite{m}}{}$ can be seen as a generalization of the probability current $j_{\twoDsite{n},\twoDsite{m}}$. We choose to call it a \emph{non-local current} because, contrary to $j_{\twoDsite{n},\twoDsite{m}}$, the new quantity $\qPlusWithoutI{}{\twoDsite{n}\twoDsite{m}}{}$ contains amplitudes and hoppings at mapped locations $\strich{\twoDsite{n}},\strich{\twoDsite{m}}$. This non-locality of $\qPlusWithoutI{}{\twoDsite{n},\twoDsite{m}}{}$ is the cause for several different properties of the currents $\qPlusWithoutI{}{\twoDsite{n},\twoDsite{m}}{}$ and $j_{\twoDsite{n},\twoDsite{m}}$, as we will see in the following. 
Firstly, let us note that $j_{\twoDsite{n},\twoDsite{m}} = 0$ whenever the two sites $\twoDsite{n},\twoDsite{m}$ are not connected (i.e. neighbouring) sites. On the contrary, $\qPlusWithoutI{}{\twoDsite{n},\twoDsite{m}}{} = 0$ is in general only true if neither $\twoDsite{n},\twoDsite{m}$ \emph{nor} $\strich{\twoDsite{n}},\strich{\twoDsite{m}}$ are neighbours. Secondly, the probability current $j_{\twoDsite{n},\twoDsite{m}}$ features the special property of sign-inversion under direction inversion. This means that the flow from $\twoDsite{m}$ to $\twoDsite{n}$ is equal to minus the flow in the reverse direction, i.e. 
\begin{equation}
j_{\twoDsite{n},\twoDsite{m}} = - j_{\twoDsite{m},\twoDsite{n}} .
\end{equation}
Non-local currents, on the contrary, do in general \emph{not} behave like this. Therefore, the non-local current from site $\twoDsite{n}$ to site $\twoDsite{m}$ is in general \emph{not} equal to minus the non-local current from $\twoDsite{m}$ to $\twoDsite{n}$. This can be seen from the following equation:
\begin{small}
	\begin{equation}
	\qPlusWithoutI{}{\twoDsite{n},\twoDsite{m}}{} = - \qPlusWithoutI{}{\twoDsite{m},\twoDsite{n}}{} - i\cdot \left( \ZS{\Delta}{}{\twoDsite{n},\twoDsite{m}} \ZS{\Psi}{*}{\twoDsite{m}} \ZS{\Psi}{}{\strich{\twoDsite{n}}} + \ZS{\Delta}{*}{\twoDsite{n},\twoDsite{m}} \ZS{\Psi}{}{\strich{\twoDsite{m}}} \ZS{\Psi}{*}{\twoDsite{n}} \right)
	\end{equation}
\end{small}%
where $\ZS{\Delta}{}{\twoDsite{n},\twoDsite{m}} = \ZS{h}{}{\twoDsite{n},\twoDsite{m}} - \ZS{h}{}{\strich{\twoDsite{n}},\strich{\twoDsite{m}}}$ and $\ZS{\Delta}{}{\twoDsite{n},\twoDsite{m}}  \ne 0$ in general. To conclude this short excursion on the properties of non-local currents, we state that although $j_{\twoDsite{n},\twoDsite{m}}$ and $\qPlusWithoutI{}{\twoDsite{n},\twoDsite{m}}{}$ are both currents, they possess in detail different behaviours and properties. A graphical representation of the above properties will be provided in the discussion of \cref{fig:qConstancySystem2} in \cref{sec:domainsOfLocalSymmetriesAndSourceTerms}.

We will now derive the non-local continuity equation. To this end, we express \cref{eq:definitionOfQn}
\begin{equation*}
i\cdot q_{\twoDsite{n}} = \sum_{\twoDsite{m} \in \neighbourOf{\strich{\twoDsite{n}}}} h_{\strich{\twoDsite{n}},\twoDsite{m}} \cjOnlySub{\Psi_{\twoDsite{n}}} \Psi_{\twoDsite{m}} - \sum_{\twoDsite{m}\in \neighbourOf{\twoDsite{n}}} h_{\twoDsite{n},\twoDsite{m}}^{*} \Psi_{\strich{\twoDsite{n}}} \cjOnlySub{\Psi_{\twoDsite{m}}}
\end{equation*}
by means of non-local currents $\qPlusWithoutI{}{\twoDsite{n},\twoDsite{m}}{}$ and arrive at
\begin{equation} \label{eq:sumNonLocalCurrentsEqualToQn}
q_{\twoDsite{n}} = \sum_{\twoDsite{m} \in \neighbourOf{\twoDsite{n}}} \qPlusWithoutI{}{\twoDsite{n},\twoDsite{m}}{} + \sum_{\substack{\strich{\twoDsite{m}} \in \neighbourOf{\strich{\twoDsite{n}}} \\ \twoDsite{m} \notin \neighbourOf{\twoDsite{n}}}} \qPlusWithoutI{}{\twoDsite{n},\twoDsite{m}}{} .
\end{equation}
It is important to emphasize here that usually the second sum contains only some terms, and sometimes it may even be empty. In the next paragraph, we will investigate this fact in more detail. Before doing that, let us show that inserting \cref{eq:sumNonLocalCurrentsEqualToQn} into \cref{eq:nonLocalCont} gives us the \emph{non-local continuity equation}
\begin{equation} \label{eq:nonLocalContWithCurrents}
\partial_{t} \sigma_{\twoDsite{n}} = \sum_{\twoDsite{m} \in \neighbourOf{\twoDsite{n}}} \qPlusWithoutI{}{\twoDsite{n},\twoDsite{m}}{} + \sum_{\substack{\strich{\twoDsite{m}} \in \neighbourOf{\strich{\twoDsite{n}}} \\ \twoDsite{m} \notin \neighbourOf{\twoDsite{n}}}} \qPlusWithoutI{}{\twoDsite{n},\twoDsite{m}}{} -i\beta_{\twoDsite{n}} \sigma_{\twoDsite{n}} .
\end{equation}

Having derived the non-local continuity equation, let us now look more closely at the sum $\sum_{\substack{\strich{\twoDsite{m}} \in \neighbourOf{\strich{\twoDsite{n}}} \\ \twoDsite{m} \notin \neighbourOf{\twoDsite{n}}}} \qPlusWithoutI{}{\twoDsite{n},\twoDsite{m}}{} $
in \cref{eq:nonLocalContWithCurrents}. Above we stated that this sum usually contains only some or even no terms, and in the following we will investigate the conditions for this to happen. As the sum runs over all sites $\twoDsite{m}$ that are mapped to neighbours of  $\strich{\twoDsite{n}}$ but are themselves \emph{not} neighbours of $\twoDsite{n}$, it is clear that this special condition is usually only fulfilled for some sites. Moreover, the sum contains \emph{no} terms if i) both $\twoDsite{n}$ and $\strich{\twoDsite{n}}$ have the same connectivity and ii) all neighbours $\neighbourOf{\twoDsite{n}}$ of site $\twoDsite{n}$ are mapped to neighbours $\neighbourOf{\strich{\twoDsite{n}}}$ of $\strich{\twoDsite{n}}$.  Concluding the above, we state that the mapping $\TS{\systemDomain}$ is the main factor determining the number of summands in $\sum_{\substack{\strich{\twoDsite{m}} \in \neighbourOf{\strich{\twoDsite{n}}} \\ \twoDsite{m} \notin \neighbourOf{\twoDsite{n}}}} \qPlusWithoutI{}{\twoDsite{n},\twoDsite{m}}{} $, and in \cref{sec:domainsOfLocalSymmetriesAndSourceTerms} we will show the implications of this. Let us close this paragraph with a short definition: As we will need the combination of conditions i) and ii) more than once in the following, we will encapsulate them into the statement that the mapping $\TS{\systemDomain}$ \enquote{keeps the connectivity of site $\twoDsite{n}$}.

In the remainder of this work we will restrict ourselves to eigenstates, i.e. we set $\ket{\Psi} = \ket{\psi^{\nu}}$ where
\begin{equation*}
\ket{\psi^{\nu}}= \sum_{\twoDsite{n}} a_{\twoDsite{n}}^{\nu} e^{-i E^{\nu} t} \ket{\twoDsite{n}} 
\end{equation*}
is the $\nu$-th eigenstate of $\hat{H}$, with $E^{\nu}$ being the energy of this eigenstate and $a_{\twoDsite{n}}^{\nu}$ being constant. However, for better readability we will omit the eigenstate index $\nu$ in the following. Note that for eigenstates we automatically have $\partial_{t} \sigma_{\twoDsite{n}} = 0$. Thus the non-local continuity equation \cref{eq:nonLocalContWithCurrents} becomes
\begin{equation} \label{eq:nonLocalKirchhoff}
0 = \sum_{\twoDsite{m} \in \neighbourOf{\twoDsite{n}}} \qPlusWithoutI{}{\twoDsite{n},\twoDsite{m}}{} + \sum_{\substack{\strich{\twoDsite{m}} \in \neighbourOf{\strich{\twoDsite{n}}} \\ \twoDsite{m} \notin \neighbourOf{\twoDsite{n}}}} \qPlusWithoutI{}{\twoDsite{n},\twoDsite{m}}{} -i\beta_{\twoDsite{n}} \sigma_{\twoDsite{n}} .
\end{equation}
which states that the sum over all outgoing non-local currents at site $\twoDsite{n}$ is equal to a source term that is proportional to the asymmetry of the on-site potential $\beta_{\twoDsite{n}} = (v_{\strich{\twoDsite{n}}} - v_{\twoDsite{n}}^{*})$. In the following, we will call \cref{eq:nonLocalKirchhoff} the \emph{non-local Kirchhoff law} at site $\twoDsite{n}$.

\subsection{Impact of Local Symmetries on Non-Local Currents} \label{sec:domainsOfLocalSymmetriesAndSourceTerms}
\begin{figure}[tbp]
	\centering
	\includegraphics[max size={\linewidth}{0.6\textheight}]{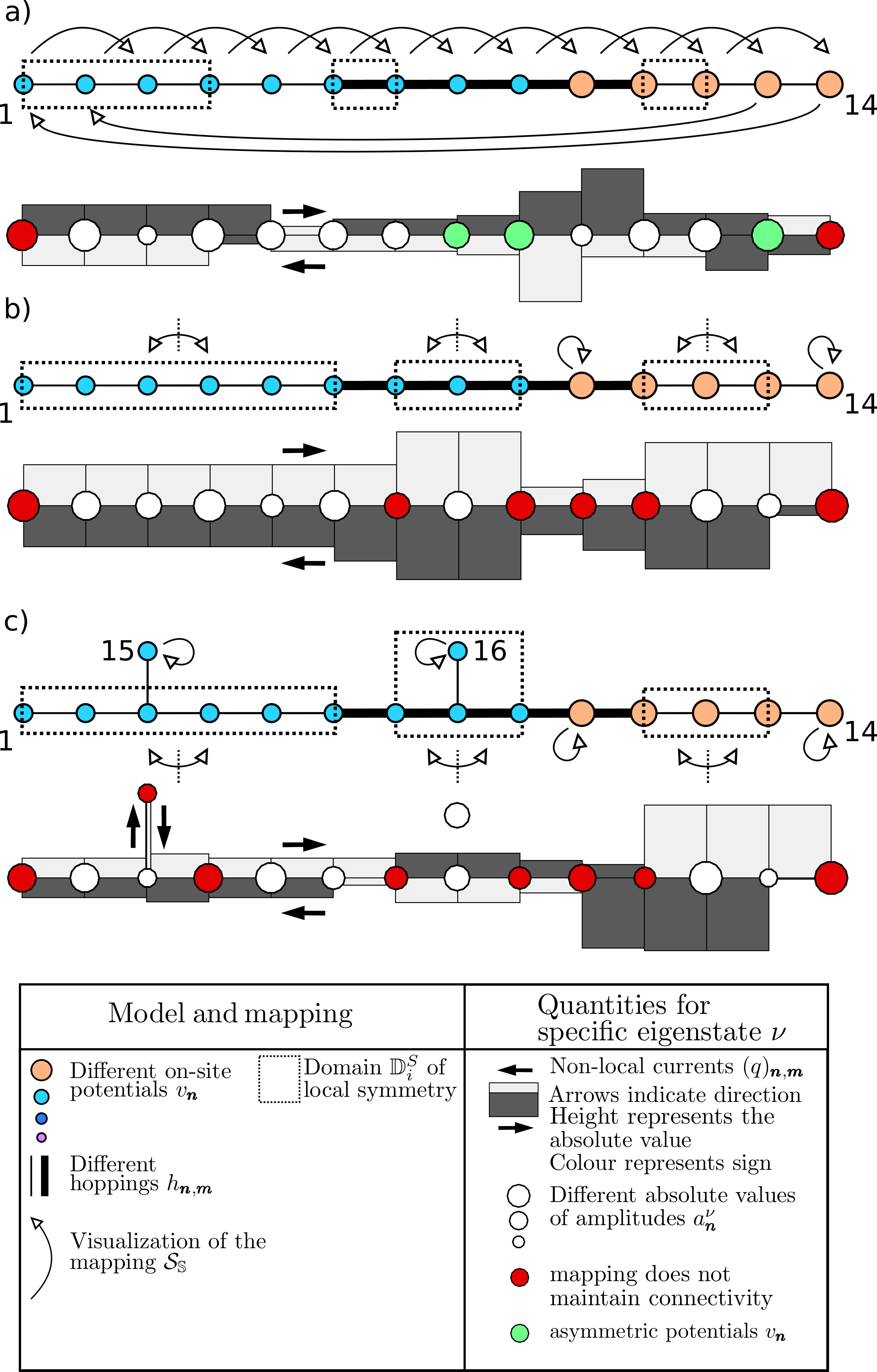}
	\caption{Visualization of non-local currents for different mappings and systems. A detailed description of this figure and of its details can be found in the text.}
	\label{fig:qConstancySystem2}
\end{figure}
So far, we have defined non-local currents and depicted the differences between the probability current $j_{\twoDsite{n},\twoDsite{m}}$ and the non-local one $\qWithoutI{}{}{\twoDsite{n},\twoDsite{m}}{}$.  It is important to stress that our formalism described above is not limited to systems possessing symmetries. In fact, until now we have not demanded any symmetries to be present at all. In this section, however, we will investigate the effects of local symmetries and graphically show the properties of the non-local currents $\qWithoutI{}{}{\twoDsite{n},\twoDsite{m}}{}$ which have been derived above. For this purpose we will make extensive use of \cref{fig:qConstancySystem2} and gradually describe and understand all of its features. The figure shows three finite setups, where the first two are 14 sites and the third setup is 16 sites in size. A model of the physical system showing the connectivity of the system as well as the hopping and on-site potential values can be found in the upper part of each subfigure. Additionally we also show the mapping $\TS{\systemDomain}$ used as well as the domains of local symmetry which are indicated by dashed boxes. In subfigure b) and c), we only indicated parts of $\TS{\systemDomain}$ by drawing two arrows and a vertical line, symbolising a local reflection mapping. In the lower part of each subfigure we show the relevant physical quantities for a chosen eigenstate. More precisely, the absolute values (different sizes) of the amplitudes as well as both absolute value and sign of non-local currents are shown. For each link between neighbouring sites, there are two non-local currents. This is necessary since, as stated above, in general $\absolute{\qWithoutI{}{}{\twoDsite{m},\twoDsite{n}}{}} \ne \absolute{\qWithoutI{}{}{\twoDsite{n},\twoDsite{m}}{}}$, and therefore both $\qWithoutI{}{}{\twoDsite{m},\twoDsite{n}}{}$ \emph{and} $\qWithoutI{}{}{\twoDsite{n},\twoDsite{m}}{}$ are needed. These two non-local currents are placed above (below) the connection between sites $\twoDsite{n}$ and $\twoDsite{m}$, corresponding to the non-local current in the right (left) direction, respectively. Their sign is color-coded, while their absolute value is represented by their height. Note that we only show non-local currents $\qWithoutI{}{}{\twoDsite{n},\twoDsite{m}}{}$ between \emph{neighbouring} sites in order to be transparent in terms of illustration.
To compensate for this omission we marked all affected sites red, i.e. we marked all sites red where the right sum $\sum_{\substack{\strich{\twoDsite{m}} \in \neighbourOf{\strich{\twoDsite{n}}} \\ \twoDsite{m} \notin \neighbourOf{\twoDsite{n}}}} \qPlusWithoutI{}{\twoDsite{n},\twoDsite{m}}{}$
in the corresponding non-local Kirchhoff equation
\begin{equation*}
0 = \sum_{\twoDsite{m} \in \neighbourOf{\twoDsite{n}}} \qPlusWithoutI{}{\twoDsite{n},\twoDsite{m}}{} + \sum_{\substack{\strich{\twoDsite{m}} \in \neighbourOf{\strich{\twoDsite{n}}} \\ \twoDsite{m} \notin \neighbourOf{\twoDsite{n}}}} \qPlusWithoutI{}{\twoDsite{n},\twoDsite{m}}{} -i\beta_{\twoDsite{n}} \sigma_{\twoDsite{n}} .
\end{equation*}
contains at least one summand, which is the case if the mapping does not maintain the connectivity of site $\twoDsite{n}$. We also marked all sites $\twoDsite{n}$ in green where $\beta_{\twoDsite{n}} = v_{\twoDsite{n}}^{*} - v_{\strich{\twoDsite{n}}} \ne 0$, i.e. where the on-site potential is not symmetric under the mapping $\TS{\systemDomain}$.

Having explained the overall structure of \cref{fig:qConstancySystem2}, let us now review our definition of a \emph{domain $\symmDomain$ of local symmetry} as done in \cref{sec:localSymmetries}. There, we defined a local symmetry in the domain $\symmDomain$ to be present if
\begin{equation} \label{eq:domainsOfLocalSymmetryDef}
H_{\twoDsite{m},\twoDsite{n}} = H_{\strich{\twoDsite{m}},\strich{\twoDsite{n}}} \forallMacro \twoDsite{m},\twoDsite{n} \in \symmDomain .
\end{equation}
In the following, we further constrain $\symmDomain$ by demanding it to be \emph{connected}. Note that this requirement does in no way restrict the possible classes of local symmetries, but, as we will see later will simplify several relevant statements and proofs.

Three aspects are important when interpreting \cref{eq:domainsOfLocalSymmetryDef}. Firstly, note that a system may possess \emph{several} domains $\symmDomain_{i}$ of local symmetry, which could even be overlapping. For \cref{fig:qConstancySystem2}, there are three domains for each setup. Secondly, the existence or non-existence of domains of local symmetries and their size not only depends on the system itself but also on the mapping $\TS{\systemDomain}$ used, as can be seen when comparing \cref{fig:qConstancySystem2} a) and b). Thus, for the description of local symmetries it is important to choose $\TS{\systemDomain}$ properly. In particular, since $\TS{\systemDomain}$ is required to be bijective, \emph{several mappings} $\TS{\systemDomain}, \TS{\systemDomain}', \TS{\systemDomain}'', \ldots$ might be required to describe all possible local symmetries of the system, especially if these symmetries are overlapping. In the latter case, the notation $\strich{\twoDsite{n}} = \T{\systemDomain}{\twoDsite{n}}$ would be ambiguous. To ease readability, with the exception of \cref{fig:qConstancySystem2}, we solely focus on a single mapping $\TS{\systemDomain}$, rendering $\strich{\twoDsite{n}}$ unambiguous. Lastly, it is important to note that our definition \cref{eq:domainsOfLocalSymmetryDef} of a domain $\symmDomain$ of local symmetry only takes \emph{internal} connections within $\symmDomain$ into account, as can be seen in the upper right domain of local symmetry in \cref{fig:localSymmetricSystem}.

A main result of the treatment of one-dimensional systems in \cite{2016arXiv160706577M} was a special property of non-local currents: Their $\symmDomain_{i}$-domainwise constancy in strictly one-dimensional systems, which is depicted in \cref{fig:qConstancySystem2}. The proof of this constancy can be found in \cite{2016arXiv160706577M} and, in order to be self-contained, in \cref{sec:ProofConstancy} of this work. Below we will provide a modified version of this statement valid for planar systems.

We are now equipped with everything needed to understand the lower part of each subfigure in \cref{fig:qConstancySystem2}. 
Let us start with the upper two figures, where each system possesses 14 sites. To simplify the notation and to ease the readability, we will set $a_{0} = a_{15} = 0 \Rightarrow \qPlusWithoutI{}{n,0}{} = \qPlusWithoutI{}{n,15}{} = 0 \forallMacro 0 \le n \le 14$ in the treatment of these two subfigures. We first concentrate on sites which are marked neither in green or red, i.e. the white ones. At these sites $\twoDsite{n}$, the non-local Kirchhoff law is given by
\begin{equation*}
0 = \sum_{\twoDsite{m} \in \neighbourOf{\twoDsite{n}}} \qPlusWithoutI{}{\twoDsite{n},\twoDsite{m}}{} .
\end{equation*}
As all of these sites have two neighbours, one to the left and one to the right, we can write it as
\begin{equation*}
0 =  \qPlusWithoutI{}{n,n+1}{} + \qPlusWithoutI{}{n,n-1}{} \Rightarrow \qPlusWithoutI{}{n,n+1}{} = - \qPlusWithoutI{}{n,n-1}{}
\end{equation*}
where we have switched from two-dimensional notation $\twoDsite{n}$ to its one-dimensional counterpart $n$.
Therefore, the two outgoing non-local currents have equal absolute values and different signs, shown also in \cref{fig:qConstancySystem2}. Note that within each domain $\symmDomain_{i}$ of local symmetry, the non-local currents are constant. This directly follows from the above if we use the identity $q_{\twoDsite{n},\twoDsite{m}} = -q_{\twoDsite{m},\twoDsite{n}}$ for $\twoDsite{n},\twoDsite{m} \in \symmDomain_{i}$. Before proceeding, let us remind the reader that $q_{\twoDsite{n},\twoDsite{m}} = -q_{\twoDsite{m},\twoDsite{n}}$ may also be valid if $\twoDsite{n},\twoDsite{m} \notin \symmDomain_{i}$. As shown in \cref{sec:oneD}, this is true if $h_{\twoDsite{n},\twoDsite{m}} = h_{\strich{\twoDsite{n}},\strich{\twoDsite{m}}}$ which is the case for sites $7,8,9$ in subfigure a). Having investigated the sites marked in white, we next look at the sites which are marked in green. Their non-local Kirchhoff law becomes
\begin{equation*}
0 =  \qPlusWithoutI{}{n,n+1}{} + \qPlusWithoutI{}{n,n-1}{} -i \beta_{n} a_{n}^{*}a_{\strich{n}} .
\end{equation*}
with $\beta_{n} = (v_{n}^{*} - v_{\strich{n}}) \ne 0$. Therefore, the absolute values of the two outgoing non-local currents are not equal. Finally, we focus on the sites marked in red. Their non-local Kirchhoff law is given by
\begin{equation*}
0 = \sum_{\twoDsite{m} \in \neighbourOf{\twoDsite{n}}} \qPlusWithoutI{}{\twoDsite{n},\twoDsite{m}}{} + \sum_{\substack{\strich{\twoDsite{m}} \in \neighbourOf{\strich{\twoDsite{n}}} \\ \twoDsite{m} \notin \neighbourOf{\twoDsite{n}}}} \qPlusWithoutI{}{\twoDsite{n},\twoDsite{m}}{} 
\end{equation*}
where the right sum contains at least one term. Therefore, there are more than two non-local currents in total and unless the second sum vanishes we have $\absolute{\qPlusWithoutI{}{n,n+1}{}} \ne \absolute{\qPlusWithoutI{}{n,n-1}{}}$. Having analysed the upper two subfigures, let us focus on the third one. Comparing it to subfigure b), one can see that the only difference are the two additional sites $15$ and $16$. At site $3$, the non-local Kirchhoff law contains an additional non-local current $q_{3,15}$, and thus we no longer have $q_{3,2} = -q_{3,4}$. At site $8$, however, we do have $q_{8,7} = -q_{8,9}$. This is due to the mapping: Since the sites $8$ and $16$ are mapped onto themselves, the currents $q_{8,16} = q_{16,8} = 0$ identically vanish. Apart from this difference, all the discussion on subfigure a) and b) also applies to subfigure c).

We are now ready to state the conditions needed to extend the concept of $\symmDomain_{i}$-domainwise constancy of non-local currents to planar systems. If one combines the properties described above, one comes to the conclusion that
non-local currents are $\symmDomain_{i}$-domainwise constant if i) all sites within $\symmDomain_{i}$ have less than three neighbours and ii) the mapping $\TS{\systemDomain}$ maintains the connectivity of all sites $\twoDsite{n}$ within $\symmDomain_{i}$. Combining conditions i) and ii) with demanding the existence of the domain $\symmDomain_{i}$ of local symmetry has two effects: Firstly, $\qPlusWithoutI{}{\twoDsite{n},\twoDsite{m}}{} = - \qPlusWithoutI{}{\twoDsite{m},\twoDsite{n}}{} \forallMacro \twoDsite{n},\twoDsite{m} \in \symmDomain_{i}$. Secondly, the non-local Kirchhoff law for a site $\twoDsite{n} \in \symmDomain_{i}$ consists of exactly one \emph{or} two non-local currents, depending on the connectivity of site $\twoDsite{n}$. Combining the above, one can easily conclude that the non-local currents within $\symmDomain_{i}$ are constant, as claimed above.

Let us conclude this section with two important statements. Firstly, note that non-local currents induce an \emph{additional structure on the eigenstates} of the system. Thus, one gains deeper insights into the nature of locally symmetric systems by studying non-local currents and their behaviour. Secondly and most importantly, already at this level our framework enables the reader to use it as a powerful tool to search and find suitable locally symmetric systems exhibiting the effects of local symmetries. To demonstrate this, note that in \cref{fig:qConstancySystem2} a) and b), all non-local currents shown are non-vanishing, and the only non-local current in subfigure c) that vanishes does so because of the trivial mapping at its endpoints. An important question then reads: What happens if a given non-local current $\qPlusWithoutI{}{\twoDsite{n},\twoDsite{m}}{}$ vanishes, given a non-trivial mapping? This question can be answered as follows: If the two sites $\twoDsite{n},\twoDsite{m}$ are elements of the same domain $\symmDomain_{i}$ of local symmetry, then we have $0 = \qPlusWithoutI{}{\twoDsite{n},\twoDsite{m}}{} = h_{\twoDsite{n},\twoDsite{m}} \left( a_{\twoDsite{n}}^{*} \ZS{a}{}{\strich{\twoDsite{m}}} -   a_{\strich{\twoDsite{n}}} a_{\twoDsite{m}}^{*} \right)$ and if $h_{\twoDsite{n},\twoDsite{m}} \ne 0$ we get the important relation
\begin{equation} \label{eq:EffectOfVanishingNLC}
	a_{\twoDsite{n}}^{*} \ZS{a}{}{\strich{\twoDsite{m}}} = a_{\strich{\twoDsite{n}}} a_{\twoDsite{m}}^{*} 
\end{equation}
which connects the four amplitudes $a_{\twoDsite{n}},a_{\strich{\twoDsite{n}}},a_{\twoDsite{m}},a_{\strich{\twoDsite{m}}}$. If additionally $a_{\strich{\twoDsite{n}}}, a_{\strich{\twoDsite{m}}} \ne 0$ we have
\begin{equation*}
	\frac{a_{\twoDsite{n}}^{*}}{\ZS{a}{}{\strich{\twoDsite{n}}}} = \frac{a_{\twoDsite{m}}^{*}}{\ZS{a}{}{\strich{\twoDsite{m}}}} .
\end{equation*}
Therefore a vanishing non-local current allows one to directly connect its constituent amplitudes. In this manner, one can clearly see the effects of local symmetries on the eigenstates of the system. This effect can be increased further: If one connects a vanishing non-local current with the $\symmDomain_{i}$-domainwise constancy of non-local currents described above, \cref{eq:EffectOfVanishingNLC} becomes valid in the \emph{whole} domain $\symmDomain_{i}$, and thus the effects of local symmetries can clearly be seen. We will present such a system in  \cref{sec:OpenEndedChains}.

\subsection{Allowed Symmetries} \label{sec:PermutationMapping}
\begin{figure}[tbp]
	\centering
\includegraphics[max size={0.3\linewidth}{0.5\textheight}]{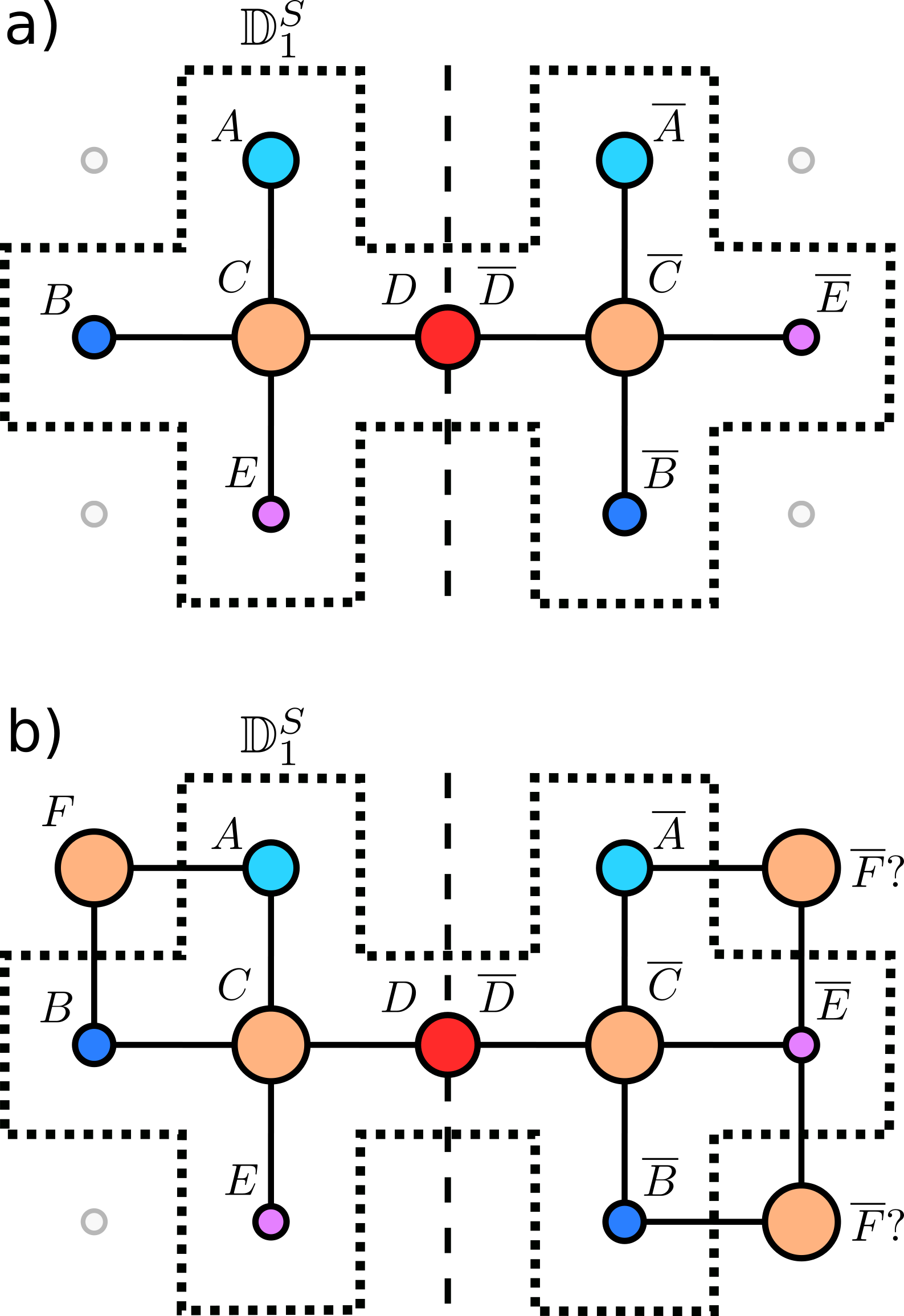}
	\caption{a): A symmetry of the second class (compare text) constructed by taking a reflection mapping (thin dashed line) as a basis and subsequently permuting $\protect\strich{B}$ and $\protect\strich{E}$. The thick dotted line delimits the domain $\protect\symmDomain_{1}$ of local symmetry. b): Already a small extension of the system reveals the limitations of symmetries of the second class. In order to add $F$ into the domain of local symmetry, the mapped sites must simultaneously fulfil $h_{A,F} = h_{\protect\strich{A},\protect\strich{F}}$ and $h_{B,F} = h_{\protect\strich{B},\protect\strich{F}}$.  This constraint is fulfilled by none of the possible positions of $\protect\strich{F}$. Therefore, the site $F$ can not be included into $\protect\symmDomain_{1}$.%
	}
	\label{fig:permutationMapping}
\end{figure}

In this section we discuss the different possibilities for local symmetries in planar discrete systems. These symmetries can be divided into two classes, with the first of them consisting of mappings which may be graphically interpreted as translation, reflection, rotation and combinations thereof. The second class of symmetries consists of more general ones that may \emph{not} be interpreted in that way. However, they can be constructed by taking a symmetry from the first class as a basis and subsequently permuting some sites. An example for this is given in \cref{fig:permutationMapping} a), where we get a symmetry of the second class by taking a reflection symmetry as an ingredient and subsequently permute the sites $\strich{B}$ and $\strich{E}$. The resulting symmetry can not be described by a combination of translation, reflection and rotation symmetries.

Symmetries of the second class come with one severe restriction: In many cases, one can not design a system with a global symmetry of that kind. To show this, let us extend our system of \cref{fig:permutationMapping} a) slightly. Namely, we add a site $F$ which represents another common neighbour of $A$ and $B$, as shown in \cref{fig:permutationMapping} b). Compared to \cref{fig:permutationMapping} a), the symmetry $\symmDomain_{1}$ is now a \emph{local} one. Can we extend it to be a global symmetry again? In order to achieve this, the site $F$ must be included into the domain $\symmDomain$ of local symmetry. This in turn requires one to find a suitable site $\strich{F}$ such that $h_{\strich{A},\strich{F}} = h_{A,F}$ and $h_{\strich{B},\strich{F}} = h_{B,F}$. But since only next-neighbour hoppings are allowed within the scope of this work, \emph{only one} of these two constraints can be fulfilled. Therefore, if neither of the two hoppings $h_{A,F}, h_{B,F}$ are zero, the domain of local symmetry can never be extended to the complete system.

\subsection{Summed currents} \label{sec:SummedCurrents}
\begin{figure}[htbp]
	\centering
	\includegraphics[max size={0.5\linewidth}{0.5\textheight}]{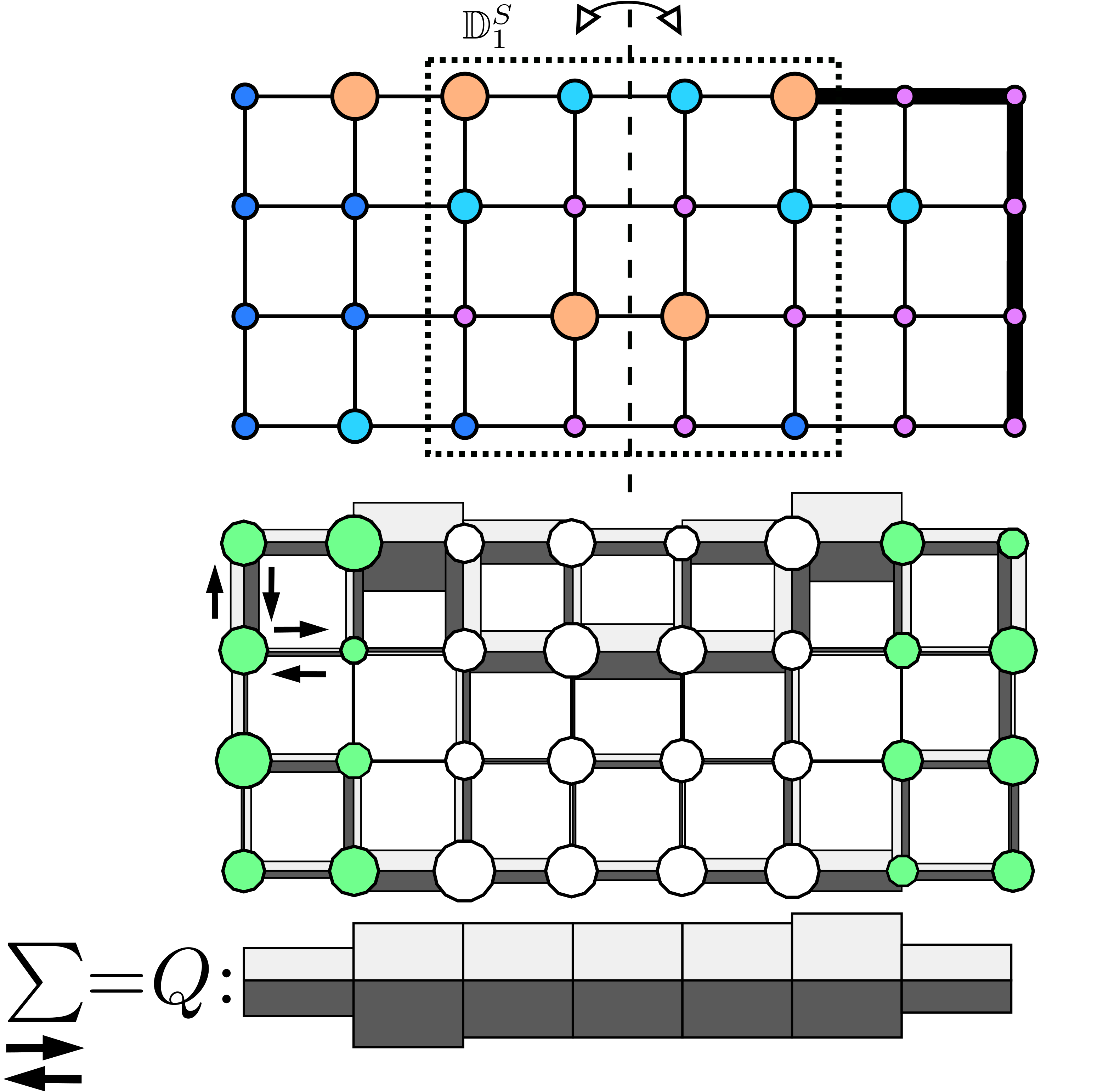}
	\caption{Visualization of the constancy of summed non-local currents $Q$ (bottom of picture). The mapping $\TS{\systemDomain}$ is a reflection about the vertical axis indicated by a dashed line. For an explanation of the used symbols, see caption of \cref{fig:qConstancySystem2}.}
	\label{fig:summedCurrents}
\end{figure}
Contrary to the one-dimensional case, non-local currents in planar systems are in general \emph{not constant} throughout a domain $\symmDomain_{i}$ of local symmetry. In the following, however, we will give conditions under which a \emph{constant} summed non-local current $Q$ can be derived by summing over individually non-constant $\qWithoutI{}{}{\twoDsite{n},\twoDsite{m}}{}$.
Before we state these conditions, we first need to define a suitable region $\mathcal{R}$ in which the conditions should apply.
We define this region $\mathcal{R}$ as a connected set of sites that contains \emph{all} sites within its boundaries, i.e. from $x_{min}$ to $x_{max}$ in x-direction and from $y_{min}(x)$ to $y_{max}(x)$ in $y$-direction. We further demand $\mathcal{R}$ to be confined by demanding $\mathcal{R}$ to have vanishing vertical coupling to the above at $(x,y_{max}(x))$ and below at $(x,y_{min}(x))$, respectively. 
The net non-local current
\begin{equation} \label{eq:summedCurrentEquation}
	Q_{x,x\pm1} = \sum_{y=1+y_{min}(x)}^{y_{max}(x)-1} \qPlusWithoutI{}{(x,y),(x\pm1,y)}{}
\end{equation}
is then constant \emph{within} $\mathcal{R}$ if i) all sites within $\mathcal{R}$ are elements of the same domain $\symmDomain_{i}$ of local symmetry and ii) the mapping $\TS{\systemDomain}$ keeps the connectivity of all sites within $\mathcal{R}$. The constancy of \cref{eq:summedCurrentEquation} is proven in \cref{sec:SummedCurrentsProof} and is visualized in \cref{fig:summedCurrents}. Note that one can easily derive a constant non-local current in $y$-direction if one slightly changes the above definitions.
\section{Case studies} \label{sec:exampleSystems}

\subsection{Locally symmetric open-ended chains} \label{sec:OpenEndedChains}
\begin{figure}[tbp]
	\centering
	\includegraphics[max size={\linewidth}{0.5\textheight}]{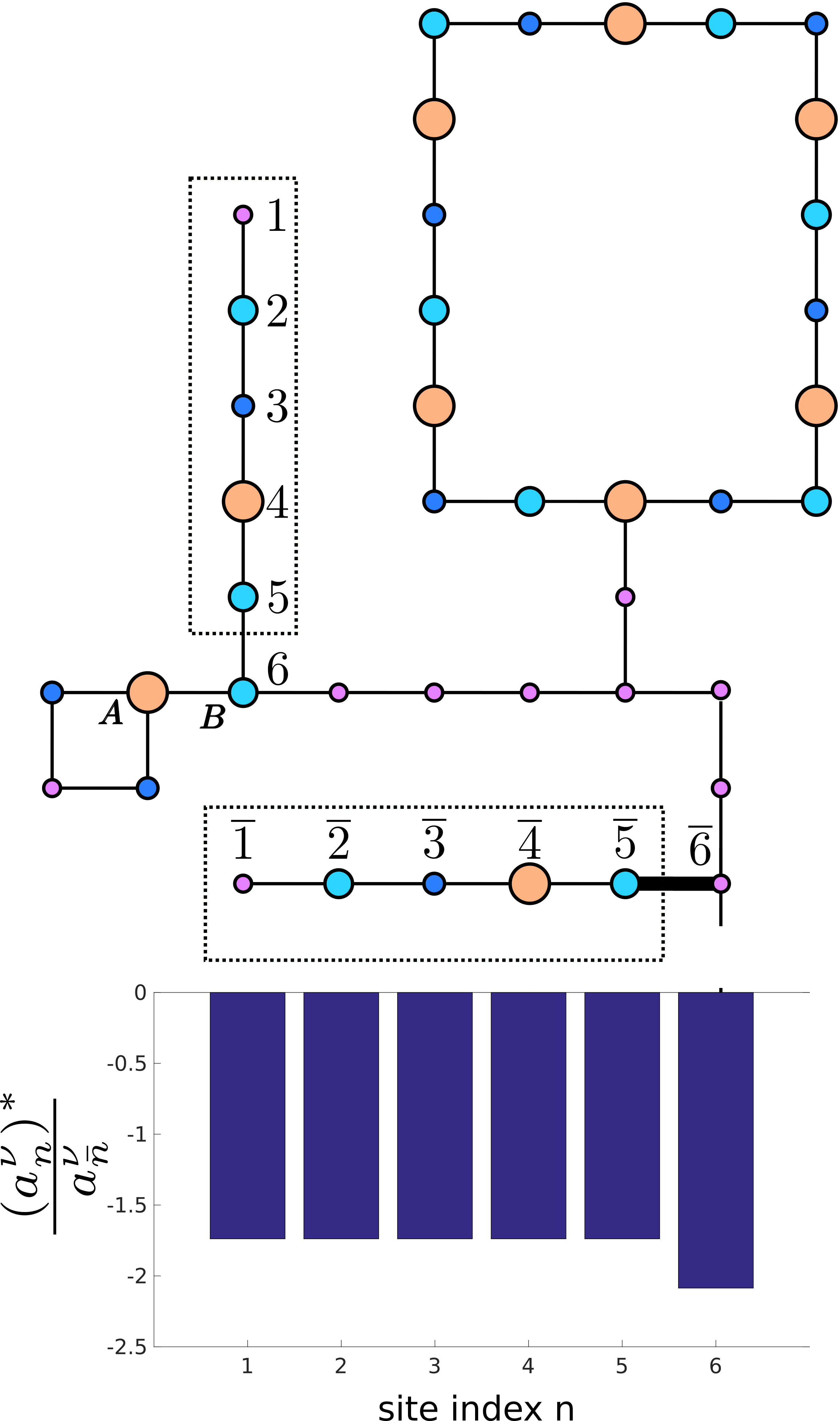}
	\caption{Top: A system possessing locally symmetric open-ended chains (surrounded by dashed boxes). Bottom3: Ratio of the amplitudes at sites $1\ldots 6$ and $\protect\strich{1}\ldots \protect\strich{6}$ for a specific eigenstate, clearly visualizing \cref{eq:openEndBoundaryReal}.}
	\label{fig:openEndedChain}
\end{figure}
In this subsection we will answer the question posed at the end of \cref{sec:domainsOfLocalSymmetriesAndSourceTerms}, namely: How could one design a subsystem in such a way that the non-local currents within it vanish? Answering this question is of high relevance and importance, as the existence of such a subsystem would clearly show the effects of its local symmetries: It would feature a constant ratio of amplitudes $a_{\twoDsite{n}}$ within the subsystem and their mapped counterparts $a_{\strich{\twoDsite{n}}}$. In other words, the subsystem's amplitudes are similar to their mapped counterparts, clearly demonstrating the influence of local symmetries on the subsystem's eigenstates.

In this section, we will present a subsystem that accomplishes this: Locally symmetric one-dimensional open-ended chains (see \cref{fig:openEndedChain}). The basic idea to achieve vanishing non-local currents is as follows: In the absence of sink terms due to asymmetries, the non-local Kirchhoff law at site $\twoDsite{n}$ reads $0 = \sum_{\twoDsite{m} \in \neighbourOf{\twoDsite{n}}} \qPlusWithoutI{}{\twoDsite{n},\twoDsite{m}}{} + \sum_{\substack{\strich{\twoDsite{m}} \in \neighbourOf{\strich{\twoDsite{n}}} \\ \twoDsite{m} \notin \neighbourOf{\twoDsite{n}}}} \qPlusWithoutI{}{\twoDsite{n},\twoDsite{m}}{}$. Provided that the right-hand side of this equation contains only \emph{one} non-local current, this current automatically vanishes.

Having presented the overall idea, let us now describe how we achieve a vanishing non-local current by using a one-dimensional open-ended chain.
A site $\twoDsite{n}=1$ located at the end of such a chain has by definition only one neighbour, i.e. $\absolute{\neighbourOf{1}} = 1$, and we denote this neighbour by site $2$. If both the mapping and the system are designed accordingly, then the non-local Kirchhoff law at site $1$ solely consists of a \emph{single} non-local current $\qPlusWithoutI{}{1,2}{}$ which, as shown above, automatically vanishes. If the open-ended chain also fulfils the conditions for the $\symmDomain_{i}$-domainwise constancy of non-local currents, then \emph{all} non-local currents within the open-ended chain do vanish.
For open-ended chains, $\symmDomain_{i}$-domainwise constancy is achieved if i) the last $1\ldots k$ (counted from the open end) of such an open-ended chain are elements of the same domain of local symmetry $\symmDomain_{i}$ and ii) the mapping $\TS{\systemDomain}$ keeps the connectivity of these $k$ sites. Note that the combination of i) and ii) is equal to demanding the system to have a second open-ended chain that is a \emph{duplicate} of the first, as can be seen in \cref{fig:openEndedChain}.

Let us now quantify the effects of the above conditions. As is be proven in \cref{sec:ProofOpenEnd}, the vanishing of non-local currents is equivalent to the fact that the ratio of the complex conjugates of the amplitude at the $k$ sites and their symmetry transformed counterparts is related by a constant. Put into equations, if we look at the $\nu$-th eigenstate, then
\begin{equation} \label{eq:openEndBoundaryReal}
\begin{pmatrix}
\cj{a_{1}^{\nu}} \\
\vdots \\
\cj{a_{k}^{\nu}} 
\end{pmatrix}
=
C^{\nu} \cdot 
\begin{pmatrix}
a_{\strich{1}}^{\nu}  \\
\vdots \\
a_{\strich{k}}^{\nu}
\end{pmatrix} 
\end{equation}
where $C^{\nu}$ is constant for each eigenstate $\nu$ and $1 \ldots , k$ are sites within an open-ended one-dimensional chain. In order for \cref{eq:openEndBoundaryReal} to be true, both of the vectors are demanded to be non-vanishing. In other words, we demand both $a_{n}^{\nu}, a_{\strich{n}}^{\nu} \ne 0 \; , \; n=1\ldots k$. A graphical visualization of \cref{eq:openEndBoundaryReal} is given in \cref{fig:openEndedChain}.
\subsection{Closed loop systems}
\begin{figure}[htbp]
	\centering
	\includegraphics[max size={0.5\linewidth}{0.5\textheight}]{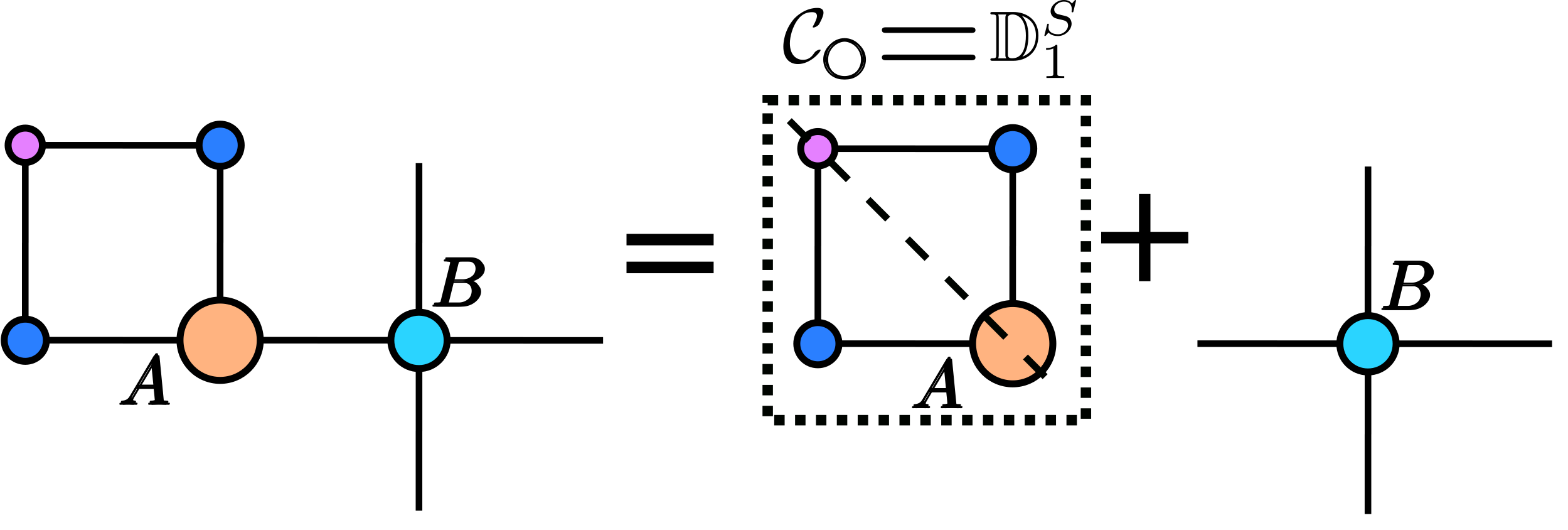}
	\caption{The splitting into $\closedLoopDomain$ and the remainder of the system. Note that in this particular case $\closedLoopDomain$ is an example of a locally reflection symmetric loop with site $\twoDsite{A}$ lying on the line of reflection (diagonal line), as mentioned in the text.}
	\label{fig:LoopAufteilung}
\end{figure}
In this section we will finally show two classes of subsystems where some or all amplitudes of their eigenstates are locally (anti)symmetric. Note that, compared to the results of \cref{sec:OpenEndedChains}, local symmetry of eigenstates is a far more restrictive property. 
The basis of our analysis are one-dimensional closed loops $\closedLoopDomain$, i.e. one-dimensional chains whose ends are connected. These closed loops may either be totally independent, or they may be coupled to a greater system. For the treatment of local symmetries, we are only interested in the second case, but in the following we will first treat the former one. This will help us seeing how the change from global to local symmetry affects the eigenstates.

Let us now assume the closed loop $\closedLoopDomain$ to be isolated. $\closedLoopDomain$ may then feature two different symmetries: Reflection and clockwise translation. Since $\closedLoopDomain$ is isolated, the symmetry is a global one, and thus one can choose the eigenstates to possess a definite parity (for reflection symmetry) or to be a Bloch function (for translation), respectively.

How does this change if we make $\closedLoopDomain$ part of an \emph{arbitrary} system $\systemDomain \supset \closedLoopDomain$? To answer this question, we add a connection between the sites $\twoDsite{A} \in \closedLoopDomain$ and $\twoDsite{B} \notin \closedLoopDomain$ (see \cref{fig:LoopAufteilung}). This connection does \emph{not} break the symmetry within $\closedLoopDomain$, but transforms the former global symmetry within $\closedLoopDomain$ into a local symmetry within $\closedLoopDomain$. As a consequence, if we let the operator $\hat{\Sigma}_{L}$ describe the reflection or translation symmetry within $\closedLoopDomain$ while it acts as the identity operator everywhere else, then in most cases this operator does no longer commute with the Hamiltonian of the complete system.

As the first of the two subsystems shown in this section, let us present a case where $[\hamilt,\hat{\Sigma}_{L}]=0$. The system that features this commutation is a closed loop which is reflection symmetric w.r.t. an axis of reflection that runs through site $\twoDsite{A}$, as shown in \cref{fig:LoopAufteilung}. Therefore, the eigenstates of $\hat{H}$ can be chosen to have definite local parity within $\closedLoopDomain$. Let us now prove that $[\hamilt,\hat{\Sigma}_{L}] = 0$ for this system. To this end, we define the mapping $\TS{\systemDomain}$ which corresponds to the operator $\hat{\Sigma}_{L}$, i.e. the mapping $\TS{\systemDomain}$ acts as a reflection within $\closedLoopDomain$ and as the identity mapping everywhere else. Note that each of the two sites $\twoDsite{A}$ and $\twoDsite{B}$ are mapped onto themselves under $\TS{\systemDomain}$, and thus one can easily prove that
\begin{equation*}
	H_{\twoDsite{m},\twoDsite{n}} = H_{\strich{\twoDsite{m}},\strich{\twoDsite{n}}} \forallMacro \twoDsite{m},\twoDsite{n} \in \systemDomain ,
\end{equation*}
i.e. the Hamiltonian is not just locally, but globally invariant under the mapping $\TS{\systemDomain}$. It is important to stress that this global invariance is \emph{independent} of how the system looks like outside of $\closedLoopDomain$, since we construct our mapping $\TS{\systemDomain}$ such that all sites outside of $\closedLoopDomain$ are mapped onto themselves. Therefore $[\hamilt,\hat{\Sigma}_{L}] = 0$ and the eigenstates of the system may be chosen to have definite \emph{local parity} within $\closedLoopDomain$, where the centre of reflection is a line through site $\twoDsite{A}$ that divides $\closedLoopDomain$ into two equally sized parts (see \cref{fig:LoopAufteilung}). This is a remarkable result, as it enables us to \enquote{decouple} the closed loop from the rest of the system in the sense that we can predict the structure of eigenstates within the loop, no matter how these eigenstates look like in the remainder of the system.

\begin{figure}[htbp]
	\centering
	\includegraphics[max size={\linewidth}{0.5\textheight}]{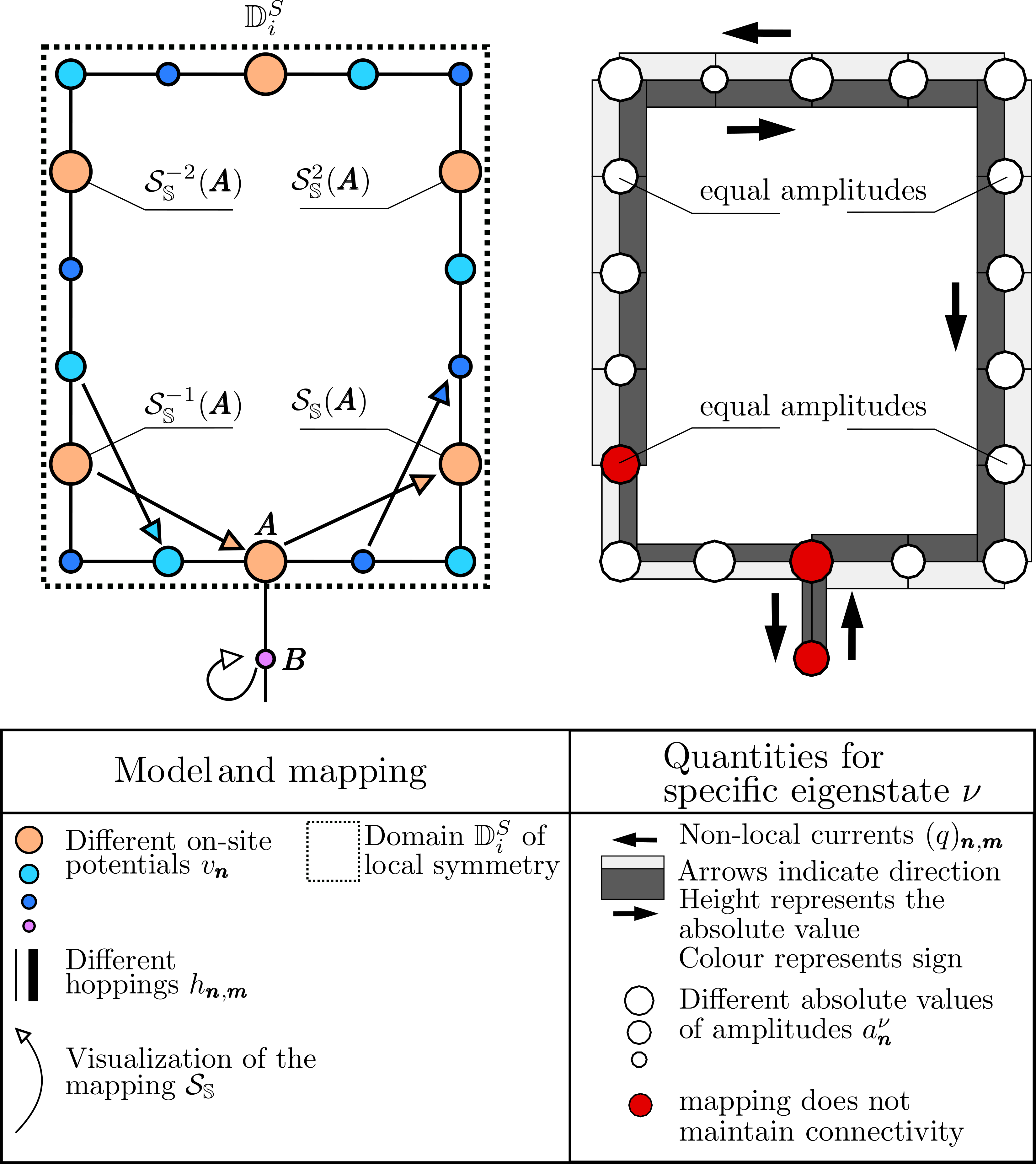}
	\caption{Left: A closed-loop system featuring an internal translational symmetry. Arrows indicate the mapping used, where colours have been added  and not all arrows have been showed to increase the readability. Right: The right-hand side shows the non-local currents for a specific eigenstate. Note the occurrence of two pairs of equal amplitudes, which represent the local symmetry of the eigenstate and whose position is given by \cref{eq:multipleEquality}.}
	\label{fig:TranslationClosedLoop}
\end{figure}
Let us now consider a second type of system. Again, we address a closed loop $\closedLoopDomain$ connected to an arbitrary system through a connection between the sites $\twoDsite{A} \in \closedLoopDomain$ and $\twoDsite{B} \notin \closedLoopDomain$. However, contrary to the former case of a local reflection symmetry, this time we assume a local clockwise translational symmetry within $\closedLoopDomain$. The operator $\hat{\Sigma}_{L}$ of local symmetry therefore describes a clockwise translation by $L$ sites within $\closedLoopDomain$ and acts as the identity operator everywhere else. As one can easily prove, the two operators $\hat{\Sigma}_{L}$ and $\hamilt$ \emph{do not commute}, contrary to the case of local reflection symmetry treated above. However, certain eigenstates of this setup can be chosen to be locally symmetric. This local symmetry becomes apparent if one chooses the eigenstates to be real-valued, which is always possible due to the restriction of real-valued hoppings. The local symmetry can then be described as
\begin{equation} \label{eq:translationClosedLoopResult}
\TInverse{\systemDomain}{\ZS{a}{\nu}{\twoDsitesoft{A}}} = \T{\systemDomain}{\ZS{a}{\nu}{\twoDsitesoft{A}}} 
\end{equation}
and is present for all eigenstates $\nu$ with $a_{\twoDsite{B}}^{\nu} \ne 0$. In the above \cref{eq:translationClosedLoopResult}, $\TS{\systemDomain}$ is the mapping which corresponds to the operator $\hat{\Sigma}_{L}$, i.e. within the closed loop it describes a clockwise translation by $L$ sites and acts as the identity mapping everywhere else. Its inverse $\TSInverse{\systemDomain}$ describes a counter-clockwise translation by $L$ sites. In this sense, the eigenstates feature a \emph{conditional local symmetry} for the two amplitudes $\TInverse{\systemDomain}{\ZS{a}{\nu}{\twoDsitesoft{A}}}$ and $\T{\systemDomain}{\ZS{a}{\nu}{\twoDsitesoft{A}}}$.
If the closed loop is big enough, \cref{eq:translationClosedLoopResult} can also be used to relate more than just two amplitudes. To show this, suppose that inside $\closedLoopDomain$ the mapping $\TS{\systemDomain}$ denotes a translation by $L$ sites. $L$ reflects the symmetry, and if the closed loop is big enough, then it is automatically also symmetric under translations of $k\cdot L$ sites where $k$ is a natural number. To describe this local symmetry, we introduce a new mapping $\TS{\systemDomain}'$ that describes a translation by $k\cdot L$ sites. By repeating the above procedure for the mapping $\TS{\systemDomain}'$ we get
\begin{equation*}
\TS{\systemDomain}^{'} (\ZS{a}{i}{\twoDsitesoft{A}}) = \TS{\systemDomain}^{'} (\ZS{a}{i}{\twoDsitesoft{A}}) .
\end{equation*}
This identity may also be written more conveniently: Since a translation by $k\cdot L$ sites may also be obtained by letting $\TS{\systemDomain}$ act $k$-times, inside the closed loop we have $\TS{\systemDomain}' = \TS{\systemDomain}^{k}$ . This yields
\begin{equation} \label{eq:multipleEquality}
\TS{\systemDomain}^{-k} (\ZS{a}{i}{\twoDsitesoft{A}}) = \TS{\systemDomain}^{k} (\ZS{a}{i}{\twoDsitesoft{A}}) .
\end{equation}
\Cref{fig:TranslationClosedLoop} visualizes a closed loop that is big enough to provide both a $L=3$ and a $L=6$ translational symmetry. Thus, we have $k=1,2$ and there are \emph{two} pairs of sites that have equal amplitudes.

\section{Conclusions}
We have extended the formalism of non-local currents in one-dimensional discrete systems \cite{2016arXiv160706577M} to models with higher site connectivity, and applied it to planar setups with local symmetries. Subsequently we have put the focus on eigenstates of the system, and for those states the current continuity leads to a non-local quantum Kirchhoff law. We showed the effects of different mappings on both the non-local currents and the description of local symmetries and incorporated the description of non-uniform connectivity into the framework.
We have also investigated two particular subsystems containing local symmetries: closed loop and open-ended chain subsystems. For both cases, the presence of either similar and locally symmetric eigenstates, respectively, are shown. By \enquote{similar} we hereby refer to an eigenstate where some of its constituent amplitudes are connected to certain other amplitudes via an eigenstate-dependent constant. These two setups show exemplarily that it is indeed possible to derive powerful conclusions about the eigenstates of a system possessing local symmetries.
\section{Acknowledgements}
Financial support by the Deutsche Forschungsgemeinschaft under the contract Schm 885/29-1 is gratefully acknowledged.
\appendix
\section{Proof of the domainwise-constancy} \label[secinapp]{sec:ProofConstancy}
In the following, we will prove the $\symmDomain_{i}$-domainwise constancy of non-local currents within one-dimensional subsystems. Let us begin with a site $\twoDsite{n}$ that lies in a domain of local symmetry, i.e. $\twoDsite{n}$ and all of its neighbours are elements of $\symmDomain_{i}$. We also demand that the mapping $\TS{\systemDomain}$ keeps the connectivity of all sites within the one-dimensional chain, i.e. an arbitrary site $\twoDsite{n}$ within the chain and its mapped counterpart $\T{\systemDomain}{\twoDsite{n}} = \strich{\twoDsite{n}}$ have to have the \emph{same number} of neighbours. Using this symmetry and the one-dimensionality, \cref{eq:nonLocalCont} reduces to the sum of only \emph{two} non-local currents:
\begin{small}
	\begin{equation} \label{eq:kirchhoffLawAtn}
	q_{c} = 0 =  \qPlusWithoutI{}{c,c-1}{} + \qPlusWithoutI{}{c,c+1}{} = - \qPlusWithoutI{}{c-1,c}{} + \qPlusWithoutI{}{c,c+1}{} .
	\end{equation}
\end{small}%
Here, we have replaced the general index $\twoDsite{n}$ that could have an arbitrary dimension by $c$ to indicate that we explicitly treat the one-dimensional case. Because the right-hand side of \cref{eq:kirchhoffLawAtn} contains only two non-local currents, these currents are equal.
Next, take one of the neighbours of $c$ and name it $c'$. Just as for $c$, the non-local Kirchhoff law at its neighbouring site $c'$ reads
\begin{equation} \label{eq:kirchhoffLawAtm}
q_{c'} = 0 = - \qPlusWithoutI{}{c'-1,c'}{} + \qPlusWithoutI{}{c',c'+1}{}
\end{equation}
and contains two non-local currents. And, just as for the site $c$, these two non-local currents are equal to each other. But because $c'$ and $c$ are neighbours, $c'$ is either equal to $c+1$ or to $c-1$, and the two sites $c'$ and $c$ share a common edge. Assuming that $c'=c+1$ without loss of generality, we can then combine the two non-local Kirchhoff laws \cref{eq:kirchhoffLawAtn,eq:kirchhoffLawAtm} to get
\begin{equation}
\qPlusWithoutI{}{c-1,c}{} = \qPlusWithoutI{}{c,c+1}{} = \qPlusWithoutI{}{c+1,c+2}{}.
\end{equation}
By induction, this pattern can be repeated. Therefore the non-local currents are constant within $\symmDomain_{i}$.

\section{Proof for Summed Currents Theorem} \label[secinapp]{sec:SummedCurrentsProof}
Let us now prove that
\begin{equation*}
Q_{x,x\pm1} = \sum_{y=1+y_{min}(x)}^{y_{max}(x)-1} \qPlusWithoutI{}{(x,y),(x\pm1,y)}{}
\end{equation*}
is constant \emph{within} a region $\mathcal{R}$ if i) all sites within $\mathcal{R}$ are elements of the same domain $\symmDomain_{i}$ of local symmetry and ii) the mapping $\TS{\systemDomain}$ keeps the connectivity of all sites within $\mathcal{R}$. To this end, we must arrive at an equation that looks like $Q_{x-1,x} = Q_{x,x+1}$. Our starting point is the non-local Kirchhoff equation at site $\twoDsite{n}$ which is given by
\begin{equation} \label{eq:qKirchhoff}
0 =\Delta_{\twoDsite{n}} +  \sum_{\twoDsite{m} \in \neighbourOf{\twoDsite{n}}} \qPlusWithoutI{}{\twoDsite{n},\twoDsite{m}}{} + \sum_{\substack{\strich{\twoDsite{m}} \in \neighbourOf{\strich{\twoDsite{n}}} \\ \twoDsite{m} \notin \neighbourOf{\twoDsite{n}}}} \qPlusWithoutI{}{\twoDsite{n},\twoDsite{m}}{}
\end{equation}
with $\Delta_{\twoDsite{n}} = (v_{\twoDsite{n}}^{*} - v_{\strich{\twoDsite{n}}}) a_{\twoDsite{n}}^{*} a_{\strich{\twoDsite{n}}}$ being the non-local source term. Next, we remove the second sum in \cref{eq:qKirchhoff}. As explained in \cref{sec:oneD} this can be done by demanding the mapping $\TS{\systemDomain}$ to keep the connectivity of site $\twoDsite{n}$. Demanding this for all sites in a given column $x$ and summing over all non-local Kirchhoff laws at this column then gives us
\begin{small}
	\begin{equation} \label{eq:sumWithVerticalqs}
	0 =  \sum_{y=1 + y_{min}(x)}^{y_{max}(x) -1} \big( \Delta_{(x,y)} + q_{(x,y)}^{\uparrow} + q_{(x,y)}^{\downarrow}  + q_{(x,y)}^{\leftarrow} +  q_{(x,y)}^{\rightarrow} \big)  .
	\end{equation}
\end{small}%
where $q_{(x,y)}^{\uparrow,\downarrow} = q_{(x,y),(x,y\pm1)}$ and $q_{(x,y)}^{\leftarrow,\rightarrow} = q_{(x,y),(x\pm1,y)}$. We now need to accomplish two steps: removing all source terms and getting rid of the vertical non-local currents $q_{x,y},(x,y\pm1)$. Since we have $\Delta_{\twoDsite{n}} = 0$ \emph{and} $q_{\twoDsite{n},\twoDsite{m}} = - q_{\twoDsite{m},\twoDsite{n}}$ within a domain $\symmDomain_{i}$ of local symmetry, this can be done by demanding all $(x,y) \in \symmDomain_{i}$ with $y_{min}(x)\le y \le y_{max}(x)$ to be elements of such a domain $\symmDomain_{i}$. The \cref{eq:sumWithVerticalqs} then simplifies to
\begin{equation*}
-Q_{x,x-1} = Q_{x,x+1}
\end{equation*}
and by demanding all sites $(x\pm1,y)$ with $y_{min}(x) \le y \le y_{max}(x)$ to be elements of the same domain $\symmDomain_{i}$ of local symmetry, then we finally get
\begin{equation*}
Q_{x-1,x} = Q_{x,x+1} .
\end{equation*}
If one repeats the above steps one can therefore derive a constant non-local current within $\mathcal{R}$.

\section{Proof for open end theorem} \label[secinapp]{sec:ProofOpenEnd}
We will prove \cref{eq:openEndBoundaryReal} which states that
\begin{equation}
\begin{pmatrix}
\cj{a_{1}^{\nu}} \\
\vdots \\
\cj{a_{k}^{\nu} }
\end{pmatrix}
=
C^{\nu} \cdot 
\begin{pmatrix}
a_{\strich{1}}^{\nu}  \\
\vdots \\
a_{\strich{k}}^{\nu}  
\end{pmatrix} 
\end{equation}
where $C^{\nu}$ is constant for each eigenstate $\nu$ and $1 \ldots , k$ (counted from the end) are sites within an open-ended one-dimensional chain. The prerequisites for this equation are a) the sites $1,\ldots, k \in \symmDomain_{i}$ are elements of the same domain of local symmetry and b) the mapping keeps the connectivity of the sites $1,\ldots,k$ and c) neither of the vectors are zero, i.e. neither $a_{1}^{\nu}=\ldots = a_{k}^{\nu} = 0$ nor $a_{\strich{1}}^{\nu}=\ldots = a_{\strich{k}}^{\nu} = 0$.

To prove \cref{eq:openEndBoundaryReal}, we start by showing the vanishing of the non-local current. This can be shown by writing the non-local Kirchhoff law for the site $1$ located at the open end (we will use the one-dimensional notation in the following, since the chain is one-dimensional anyway) which reads
\begin{equation}
0 = q_{1,2}^{\nu} .
\end{equation}
Since site $1$ is located at the open end of the chain, is has only one neighbour. Since we demanded site $1$ to lie in a domain of local symmetry and the mapping to keep the geometry of sites $1\ldots k$, there are no sink/source terms and only one non-local current present in the non-local Kirchhoff law for site $1$. Because of the constancy of $q$, this gives us $q=0$ within the locally symmetric open ended chain.

We now write one equation for each non-local current inside the chain (remember that these currents vanish) and arrive at a system of equations:
\begin{align} \label{eq:setOfZeroQs}
\cj{a_{1}^{\nu}} a_{\strich{2}}^{\nu} &= a_{\strich{1}}^{\nu} \cj{a_{2}^{\nu}}  \\
\cj{a_{2}^{\nu}} a_{\strich{3}}^{\nu} &= a_{\strich{2}}^{\nu} \cj{a_{3}^{\nu}} \nonumber \\
&\vdots \nonumber \\
\cj{a_{k-1}^{\nu}} a_{\strich{k}}^{\nu} &= a_{\strich{k-1}}^{\nu} \cj{a_{k}^{\nu}} \nonumber .
\end{align}
In the above, the sites $c\ldots k \in \symmDomain_{i}$. The hopping terms cancelled out since we are within the domain of local symmetry and all hoppings are taken to be real-valued.
In the following, let us first assume that all amplitudes occurring in the above equations are non-zero.
One may then bring \cref{eq:setOfZeroQs} to the form
\begin{equation}
\frac{\cj{a_{1}^{\nu}}}{a_{\strich{1}}^{\nu}} = \frac{\cj{a_{2}^{\nu}}}{a_{\strich{2}}^{\nu}} = \ldots = \frac{\cj{a_{k}^{\nu}}}{a_{\strich{k}}^{\nu}}
\end{equation}
which proves \cref{eq:openEndBoundaryReal}.

For the case that some amplitudes occurring in \cref{eq:setOfZeroQs} vanish, more work is needed to prove \cref{eq:openEndBoundaryReal}. For example, if $a_{z}^{\nu} = 0$ ($z<k$) while $a_{n}^{\nu} \ne 0$ for $n=1,\ldots,z-1,z+1,\ldots k$, \cref{eq:setOfZeroQs} gives us \emph{two} sets of equations:
\begin{align*}
\frac{\cj{a_{1}^{\nu}}}{a_{\strich{1}}^{\nu}} &= \ldots = \frac{\cj{a_{z-1}^{\nu}}}{a_{\strich{z-1}}^{\nu}} \\
\frac{\cj{a_{z+1}^{\nu}}}{a_{\strich{z+1}}^{\nu}} &= \ldots = \frac{\cj{a_{k}^{\nu}}}{a_{\strich{k}}^{\nu}} .
\end{align*}
In order to prove \cref{eq:openEndBoundaryReal}, we must show that a) from $a_{z}^{\nu} = 0 \Rightarrow a_{\strich{z}}^{\nu} = 0$ and b) $\frac{\cj{a_{z-1}^{\nu}}}{a_{\strich{z-1}}^{\nu}} = \frac{\cj{a_{z+1}^{\nu}}}{a_{\strich{z+1}}^{\nu}}$. Note that it directly follows from the stationary Schrödinger equation
\begin{equation} \label{eq:stationarySchroedingerEq}
0 = (-\beta + v_{n})a_{n}^{\nu} + \sum_{N(n)} h_{n,N(n)} a_{N(n)}^{\nu}
\end{equation}
that if $a_{1}^{\nu} = 0$ or if two neighbouring sites within the one-dimensional chain have vanishing amplitudes, then all amplitudes within the chain vanish. Therefore, we can restrict ourselves to the case of an isolated vanishing amplitude, i.e. $a_{z} = 0, \, a_{z\pm1} \ne 0$ where $2 < z < k-1$. From the $(z-1)$th equation in \cref{eq:setOfZeroQs},
\begin{equation}
\cj{a_{z-1}^{\nu}} a_{\strich{z}}^{\nu} = a_{\strich{z-1}}^{\nu} \cj{a_{z}^{\nu}} \nonumber ,
\end{equation}
we directly read of that $a_{\strich{z}} = 0$ and have thus proved claim a). From the stationary Schrödinger equation \cref{eq:stationarySchroedingerEq} evaluated at sites $z$ and $\strich{z}$ we get
\begin{align*}
h_{z,z-1} a_{z-1}^{\nu} &= - h_{z,z+1} a_{z+1}^{\nu} \\
h_{\strich{z},\strich{z-1}} a_{\strich{z-1}}^{\nu} &= - h_{\strich{z},\strich{z+1}} a_{\strich{z+1}}^{\nu}
\end{align*}
and dividing the complex conjugate of the first through the second equation and using the real-valuedness of the hopping terms and their local symmetry properties, we get that
\begin{equation*}
\frac{\cj{a_{z-1}^{\nu}}}{a_{\strich{z-1}}^{\nu}} = \frac{\cj{a_{z+1}^{\nu}}}{a_{\strich{z+1}}^{\nu}}
\end{equation*}
which proves claim b).

\section{Proof for the closed loop theorem} \label[secinapp]{sec:ProofClosedLoop}
To prove \cref{eq:translationClosedLoopResult}, let us start by noting that all sites within the closed loop are elements of the same domain of local symmetry $\symmDomain_{i}$. This means that $\qPlusWithoutI{}{\twoDsite{n},\twoDsite{m}}{} = - \qPlusWithoutI{}{\twoDsite{m},\twoDsite{n}}{} \forallMacro \twoDsite{m},\twoDsite{n} \in \closedLoopDomain$. Next, note that only two non-local Kirchhoff laws within $\closedLoopDomain$ contain more than two non-local currents: The non-local Kirchhoff law at site $\twoDsite{A}$, which additionally contains a third non-local current $\qPlusWithoutI{}{\twoDsite{A},\twoDsite{B}}{}$ and the non-local Kirchhoff law at site $\TInverse{\systemDomain}{\twoDsite{A}} = \twoDsite{C}$ which contains a geometric sink/source term. This term comes from the different connectivity of $\twoDsite{C}$ and $\twoDsite{A}$. 
These two non-local Kirchhoff laws read
\begin{align}
0&= \qPlusWithoutI{}{\twoDsite{A},\twoDsite{A}_{+}}{} + \qPlusWithoutI{}{\twoDsite{A},\twoDsite{A}_{-}}{} - \ZS{h}{*}{\twoDsite{A},\twoDsite{B}} a_{\strich{\twoDsite{A}}}^{*} a_{\twoDsite{B}} \label{eq:closedLoopOne} \\
0&= \qPlusWithoutI{}{\twoDsite{C},\twoDsite{C}_{+}}{} + \qPlusWithoutI{}{\twoDsite{C},\twoDsite{C}_{-}}{}  + \ZS{h}{}{\twoDsite{B},\twoDsite{A}} a_{\twoDsite{C}} a_{\twoDsite{B}}^{*} \label{eq:closedLoopTwo}.
\end{align}
Here, the terms $\twoDsite{A}_{\pm}, \twoDsite{C}_{\pm}$ denote the left/right (clockwise) neighbour of $\twoDsite{A},\twoDsite{C}$, respectively. Due to the one-dimensional and locally symmetric character of $\closedLoopDomain$, the non-local currents occurring in the above two non-local Kirchhoff laws are related by $\qPlusWithoutI{}{\twoDsite{A},\twoDsite{A}_{\pm}}{} =  -\qPlusWithoutI{}{\twoDsite{A},\twoDsite{C}_{\mp}}{}$. If we include the formerly omitted eigenstate index $\nu$ and add \cref{eq:closedLoopOne} to \cref{eq:closedLoopTwo} we thus get
\begin{equation*}
\cj{a_{\strich{\twoDsite{A}}}^{\nu}} a_{\twoDsite{B}}^{\nu} = a_{\twoDsite{C}}^{\nu} \cj{a_{\twoDsite{B}}^{\nu}} .
\end{equation*}
If we insert the definition of $\twoDsite{C}$ and choose the eigenstate $\nu$ to be real-valued (which is always possible due to the absence of complex-valued hoppings in this work), we get
\begin{equation*}
\T{\systemDomain}{a_{\twoDsite{A}}^{\nu}} a_{\twoDsite{B}}^{\nu} = a_{\TInverse{\systemDomain}{\twoDsite{A}}}^{\nu} a_{\twoDsite{B}}^{\nu} .
\end{equation*}
By further assuming that $a_{\twoDsite{B}} \ne 0$, we get $\TInverse{\systemDomain}{a_{\twoDsite{A}}^{\nu}} = \T{\systemDomain}{a_{\twoDsite{A}}^{\nu}}$ where $\T{\systemDomain}{a_{\twoDsite{A}}^{\nu}} = a_{\T{\systemDomain}{\twoDsite{A}}}^{\nu}$ and have thus proven \cref{eq:translationClosedLoopResult}.
\bibliography{NonLocalCurrentsAnd}
\end{document}